\documentclass[sigconf]{acmart}

\AtBeginDocument{%
  }

\setcopyright{acmlicensed}
\copyrightyear{2025}
\acmYear{2025}
\acmDOI{XXXXXXX.XXXXXXX}
\usepackage[linguistics]{forest}
\usepackage{algorithm}
\usepackage{algpseudocode}
\usepackage{subcaption}

\algblock{Input}{EndInput}
\algnotext{EndInput}
\algblock{Output}{EndOutput}
\algnotext{EndOutput}
\newcommand{\Desc}[2]{\State \makebox[2em][l]{#1}#2}

\acmConference[EDBT 25]{Make sure to enter the correct
  conference title from your rights confirmation emai}{March 25-28,
  2025}{Barcelona, Spain}
\acmISBN{978-1-4503-XXXX-X/18/06}

\begin{document}

\title{A Novel Framework Using Deep Reinforcement Learning for Join Order Selection}

\author{Chang Liu}

\affiliation{%
  \institution{University of Ottawa}
  \city{Ottawa}
  \state{Ontario}
  \country{Canada}
}
\email{cliu162@uottawa.ca}

\author{Amin Kamali}
\affiliation{%
  \institution{University of Ottawa}
  \city{Ottawa}
  \state{Ontario}
  \country{Canada}
}
\email{skama043@uottawa.ca}

\author{Verena Kantere}
\affiliation{%
  \institution{University of Ottawa}
  \city{Ottawa}
  \state{Ontario}
  \country{Canada}
}
\email{vkantere@uottawa.ca}

\author{Calisto Zuzarte}
\affiliation{%
  \institution{IBM Canada Lab}
  \state{Ontario}
  \country{Canada}
}
\email{calisto@ca.ibm.com}

\author{Vincent Corvinelli}
\affiliation{%
  \institution{IBM Canada Lab}
  \state{Ontario}
  \country{Canada}
}
\email{vcorvine@ca.ibm.com}
\renewcommand{\shortauthors}{Chang et al.}

\begin{abstract}
Join order selection is a sub-field of query optimization that aims to find the optimal join order for an SQL query with the minimum cost. The challenge lies in the exponentially growing search space as the number of tables increases, making exhaustive enumeration impractical. Traditional optimizers use static heuristics to prune the search space, but they often fail to adapt to changes or improve based on feedback from the DBMS. Recent research addresses these limitations with Deep Reinforcement Learning (DRL), allowing models to use feedback to dynamically search for better join orders and enhance performance over time. Existing research primarily focuses on capturing join order sequences and their representations at various levels, with limited comparative analysis of reinforcement learning methods. In this paper, we propose GTDD, a novel framework that integrates \underline{G}raph Neural Networks (GNN), \underline{T}ree-structured Long Short-Term Memory (Tree-LSTM), and \underline{D}ueling-\underline{D}QN. We conduct a series of experiments that demonstrate a clear advantage of GTDD over state-of the-art techniques.
\end{abstract}





\maketitle

\section{Introduction}
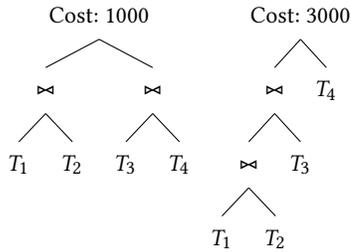
\begin{figure}[t]
\centering
\begin{forest}
  [Cost: 1000
    [$\bowtie$
     [$T_1$
     ]
     [$T_2$
     ]
    ]
    [$\bowtie$
    [
    $T_3$
    ]
    [$T_4$]
    ]
  ]
\end{forest}
\begin{forest}
[Cost: 3000
  [$\bowtie$
    [$\bowtie$
     [$T_1$
     ]
     [$T_2$
     ]
    ]
    [$T_3$]
  ]
  [$T_4$]
  ]
\end{forest}
\caption{Example of Join Order Selection for a SQL Query}
\label{fig_example}
\Description{An example of two join trees that produce the same result but have different costs.}
\end{figure}

Join order selection refers to the process of determining the sequence in which tables should be joined in an SQL query to minimize the query's execution time. The goal is to reduce the amount of data processed at each step, thereby improving overall performance. Figure \ref{fig_example} demonstrates an example of join order selection for a query with four tables, $T_1, T_2, T_3, T_4$. Even if the results of two different plan trees are identical, the cost can vary significantly depending on the join order. Traditional methods typically combine heuristic pruning strategies based on cardinality estimators and cost models to limit the search space. These strategies aim to identify and prioritize the most promising join orders, thereby reducing the computational burden while maintaining or enhancing the efficiency and performance of the query optimization process. Dynamic programming (DP)~\cite{ref_dp_based} is a practical solution that often selects the optimal plan, but tends to be resource-intensive. The problem is computationally intensive due to the factorial growth of possible join orders as the number of tables increases, resulting in higher resource consumption.

Recent research leverages Deep Reinforcement Learning (DRL) to address the limitations of traditional plan enumerators. DRL is a type of machine learning where an agent learns to make decisions by performing certain actions in an environment to achieve maximum cumulative reward. Unlike dynamic programming, DRL-based methods consider both the immediate reward of the current sub-plan and the long-term reward of the completed plan. The model utilizes feedback to search optimal join orders and avoid suboptimal ones. This approach allows the DRL agent to learn from trial and error, continuously improving the quality of the generated join orders through iterative refinement. ReJOIN~\cite{ref_ReJoin}, DQ~\cite{ref_DQ}, RTOS~\cite{ref_RTOS} and JOGGER~\cite{ref_Jogger} have demonstrated promising results, indicating that their learned optimizers can generate plans comparable to those produced by DBMS. However, DRL-based methods face two significant barriers. The first is that reinforcement learning requires information in the form of input data that accurately reflects the environment and the tasks the agent is expected to perform; The second is the inefficiency of the model, since it usually requires a large number of interactions with the environment to learn effective policies. Current research in join order selection focuses on different representation learning methods to capture representative information and not on the employed DRL method. DQ~\cite{ref_DQ}, RTOS~\cite{ref_RTOS} and JOGGER~\cite{ref_Jogger} adopt  the Deep Q Network (DQN)~\cite{ref_DQN}. While DQN represents a classical approach in DRL, there is potential for further enhancement. 

In order to tackle the aforementioned issues of DRL-based methods, we propose GTDD, a novel framework that uses Tree-LSTM~\cite{ref_treelstm} and GNN~\cite{ref_gnn} to capture an informative representation. GTDD replaces classical DQN with dueling-DQN~\cite{ref_DuelingDQN} and, furthermore, deploys curriculum learning to accelerate and stabilize the training phase.

Our main contributions in this paper, encapsulated in the GTDD framework, are the following:
\begin{itemize}
    \item We deploy various representation learning neural networks to capture different levels of representation. We use Tree-LSTM~\cite{ref_treelstm} to capture the join order sequence, GNN~\cite{ref_gnn} to capture the query representation, and finally, a multi-layer perceptron (MLP) to observe the final state representation.
    \item We propose a new reward function that compresses the reward into a ratio, given that the feedback range of disparate queries may vary.
    \item We employ the Dueling-DQN architecture to mitigate the underestimation bias inherent in vanilla DQN. This approach not only facilitates and stabilizes the training process but also enhances overall efficacy, contributing to early convergence.
    \item We conduct an extensive experimental study to evaluate the effectiveness of our proposed approach. We show that GTDD outperforms the state-of-the-art techniques.
\end{itemize}

The remainder of this paper is organized as follows: Section 2 presents the preliminaries, covering essential concepts. Related work on join order selection is reviewed in Section 3, focusing on existing solutions. Section 4 describes the problem in detail, providing a formal definition and highlighting the challenges. Section 5 is an overview of the GTDD framework. Sections 6 and 7 describe the implementation of representation learning and reinforcement learning respectively. The results are presented in Section 8 where the performance of our approach in comparison to other baselines is discussed. Finally, Section 9 concludes the paper.

\section{Preliminaries}
In order to facilitate the comprehension of our approach, we discuss preliminary concepts and techniques, necessary for the comprehension of the proposed GTDD framework.

\subsection{Query Optimization}
Query Optimization is a critical process in DBMS that focuses on improving the performance of SQL queries. A query that is  input to the DBMS undergoes an optimization process with several phases, before it is sent for execution. Query optimization aims at reducing the time of query execution. During quey optimization, the query is transformed into a set of query plans. These plans are usually represented as tree structures. Different query plans for the same SQL return the same output, but have execution times that may vary greatly. Therefore, selecting the optimal plan can reduce response time, minimize resource consumption, and lead to effectively handle large data sets, greatly improving the user experience. 

A critical part of optimal plan selection is join order selection, which determines the most efficient sequence to execute joins of tables in the query. Basic concepts in the problem of join order selection are the \emph{join plan}, \emph{partial plan} and \emph{selectivity}, which are  outlined in the following.

\paragraph{Join plan and partial plan} A join plan is a tree, where each internal node represents a join operation between base tables or intermediate results, while leaf nodes represent base tables. Given a query that includes joins, we refer to a join plan partial plan corresponds to a plan that does not include all the tables involved in the query.

\paragraph{Selectivity} Selectivity is a measure related to the number of rows returned after applying an operator. It is usually defined as the fraction of rows returned after applying a selection predicate on a table. Furthermore, \emph{join selectivity} refers to the selectivity of a join between two tables, $A$ and $B$, with respect to $A$/$B$, and is defined as the fraction of rows of $A$/$B$ that satisfy the join condition.

\subsection{Deep Reinforcement Learning (DRL)}
Reinforcement learning (RL) is a type of machine learning where an agent learns to make decisions by interacting with its environment. The agent takes actions, and based on those actions, it receives feedback in the form of rewards or penalties. The goal of the agent is to learn a policy—a strategy of actions—that maximizes the total reward over time. Q-learning~\cite{ref_qlearning} is the most classical RL algorithm that aims to learn the optimal policy in an environment. It typically maintains a q-table to store the value for all state-action pairs, and the size of the q-table would be $|A| \times |S|$ where $|A|$ is the size of action space and $|S|$ is the size of state space. For each $s\in S$, q-learning enumerates all possible actions with a q-value, and the agent selects the action with the highest q-value. As the agent explores the environment, it eventually learns a policy function $\pi(s)$ that saves the optimal actions for each state, enabling the agent to achieve its goal efficiently under the given environment. However, maintaining a Q-table becomes impractical when the state and action spaces are large. To address this challenge, Deep Reinforcement Learning (DRL)~\cite{ref_DQN} enhances RL by integrating deep neural networks, allowing the agent to handle complex environments by automatically learning important features from input data, like a join tree. The agent uses this learned knowledge to take actions that maximize the total reward over time. In Section IV, We map the join order selection problem into DRL components.
\subsection{Curriculum Learning}
The networks exhibit sensitivity to parameter changes, which adversely affect convergence and the identification of optimal join orders. Rather than training the DQN agent on randomly sampled data from the training set, a curriculum-based approach could be employed. In this approach, training queries are sampled from a sequence of increasingly complex or challenging queries. Curriculum learning~\cite{ref_curriculumlearning}, a field of human education, is a learning approach that progresses from simpler to more challenging tasks. The critical aspect is the curriculum setting. In the context of the join order selection problem, curriculum setting refers to the design of the data partition. We consider the number of participating tables in a query as the learning difficulty, where a larger number of tables in a query signifies a larger, more complex search space compared to queries with fewer tables. However, the ultimate goal is to resolve the challenging queries, which require a greater number of episodes to identify optimal join orders. Consequently, it is preferable to allocate a larger proportion of episodes to more challenging queries, rather than taking too many episodes for the simple query. The training data is initially sorted based on the number of joins and then equally split into three partitions, designated as $\{p1,\ p2,\ p3\}$, with increasing difficulty. The input query samples from these partitions. Initially, the training data samples from partition $p1$, which contains query with 3 to 5 joins. After a certain number of episodes, the training data samples from combined partition $p1 \cup p2$, which include queries from 3 to 5 joins as well as queries from 5 to 10 joins and so on. This approach is based on the rationale that learning to join orders of simple queries is easier than complex queries. Therefore, prior knowledge of the problem is beneficial to the model, allowing it to identify optimal join orders more efficiently. The process illustrates in Algorithm \ref{alg:curriculumlearning}:

\begin{algorithm}[t]
    \caption{Curriculum Learning}\label{alg:curriculumlearning}
    \begin{algorithmic}
    \Input
    \Desc{Training Set $\mathcal{Q}$ = \{$q_1,q_2,...,q_n$\}}
    \Desc{number of partition $k$}
    \Desc{updating interval $I$}
    \EndInput
    \Output
    \Desc{Updated Training Dataset $\mathcal{D}$}
    \EndOutput
    \end{algorithmic}
    \begin{algorithmic}[1]
        \State Sort the training dataset based on the number of tables participating joins
        \State split the data into $k$ partitions \{$\mathcal{P}_1,\mathcal{P}_2,...,\mathcal{P}_k$\}
        \State Initialize the empty training set $\mathcal{D}=\emptyset, i=0$
        \For{ $t= 1,2,...,T$}
            \If{$t$ is divisible by $I$ and i$<$k}
                \State $\mathcal{D} = \mathcal{D}\cup \mathcal{P}_i$
                \State $i \mathrel{+}= 1$
            \EndIf
        \EndFor
        \State \Return $\mathcal{D}$
    \end{algorithmic}
    \end{algorithm}
    

\section{Related Work}
Researchers have already proposed techniques that employ machine learning in order to tackle the join order selection problem. Furthermore, some of the proposed techniques have considered leveraging reinforcement learning. However, there has been no focus on the appropriateness of the variations of reinforcement learning for the specific problem. Additionally, the proposed techniques have not focused on the appropriateness of the representation of the input data with respect to how accurately they capture the possible actions for the construction of the full join plans from partial plans, as well as how accurately they capture the differences among plans. Nevertheless, 
informative representation of the plans and their construction can significantly improve learning efficiency. We discuss the related work from these two perspectives.


The state-of-the-art techniques that use machine learning to tackle the problem of join order selection are: DQ~\cite{ref_DQ}, Rejoin~\cite{ref_ReJoin}, RTOS~\cite{ref_RTOS} and JOGGER~\cite{ref_Jogger}. ReJoin~\cite{ref_ReJoin} and DQ~\cite{ref_DQ} use naive multilayer perceptron as their neural network model to learn the join order, and they both use vector representation to capture the state information. DQ~\cite{ref_DQ} only captures the join predicate, but ignores the underlying column and table information (e.g. selectivity), ReJoin~\cite{ref_ReJoin} takes the same approach but takes the selection predicate into account. Even though DQ and ReJOIN demonstrate some effectiveness in tackling the join order selection problem, they also exhibit a significant drawback in their state representations, as in both techniques two very distinct join trees could result in a very similar or even the same representation, due to simplistic vector representation. Such similarity misleads the model into treating different join orders as identical, causing ambiguity and leading to sub-optimal join ordering selection results.

Instead of using single vector representation and fully connected layers, RTOS~\cite{ref_RTOS} takes advantage of Tree-LSTM which reads a tree structure as input and outputs a representation of the tree, which perfectly fits the structure of the join plan. JOGGER~\cite{ref_Jogger} proposes a self-attention tree to reduce the complexity of the plan representation, and this approach has fewer parameters to learn than Tree-LSTM, which facilitates efficiency. Additionally, JOGGER was the first to highlight that training stability and convergence time could be further improved. Their work implemented curriculum learning~\cite{ref_curriculumlearning}, which partitions the training data based on difficulty (i.e., the number of joins in the query) and trains the agent with batches of data with progressively increasing complexity. While these works aim at capturing suitable state representations, yet none of them seek to improve performance from reinforcement learning perspective. 

The GTDD framework that we propose leverages experience from the discussed works and aims at filling in the gap that they leave: it holistically tackles the join order selection problem by including optimal solutions for both the representation of input data and the employment of reinforcement learning. Specifically, GTDD innovates by including various representation learning  
neural networks, which achieve accurate representation of input data at different levels.
Moreover, it proposes a novel reward function, which optimizes the selection of actions for the constructions of plans. Finally, for the first time, GTDD employs  Dueling-DQN architecture to mitigate underestimation bias in previous techniques.

Genetic algorithms (GAs) have also been explored for optimizing join orders due to their ability to efficiently search large solution spaces. GAs are particularly suitable for join order selection problems as they mimic the process of natural selection, allowing for iterative improvement through operations like mutation, crossover and selection. In the context of join order selection, each "individual" in the population represents a possible query execution plan, and the fitness function evaluates the cost of executing that plan. PostgreSQL~\cite{ref_postgres} was the first to propose a genetic optimizer prototype, but PostgreSQL deprecated the mutation operations, thus the search space is limited to the properties appearing in the initial population, which fails to handle large join queries. Later, many genetic approaches were proposed~\cite{ref_GA1,ref_GA7,ref_GA8}; however, these approaches take limited information into account and often struggle with generating valid plans. Additionally, there is extra overhead involved in fixing the plans. Furthermore, CGO~\cite{ref_CGO} proposed a more advanced genetic algorithm that is able to handle large join queries as it enables genetic operations such as crossover and mutation, which allow for diverse exploration of the search space. Moreover, CGO is specifically designed to optimize cyclic query graphs by merging the join predicates, allowing it to manage more complex query structures that other genetic optimizers struggle with. However, CGO has limitations, as it takes longer to deliver results compared to classical optimizers. Nevertheless, ML-based techniques that target join order selection, such as GTDD, prove to be superior to techniques that employ GAs.

\section{Problem Discussion}
\begin{figure}[t]
\centerline{\includegraphics[scale=0.55]{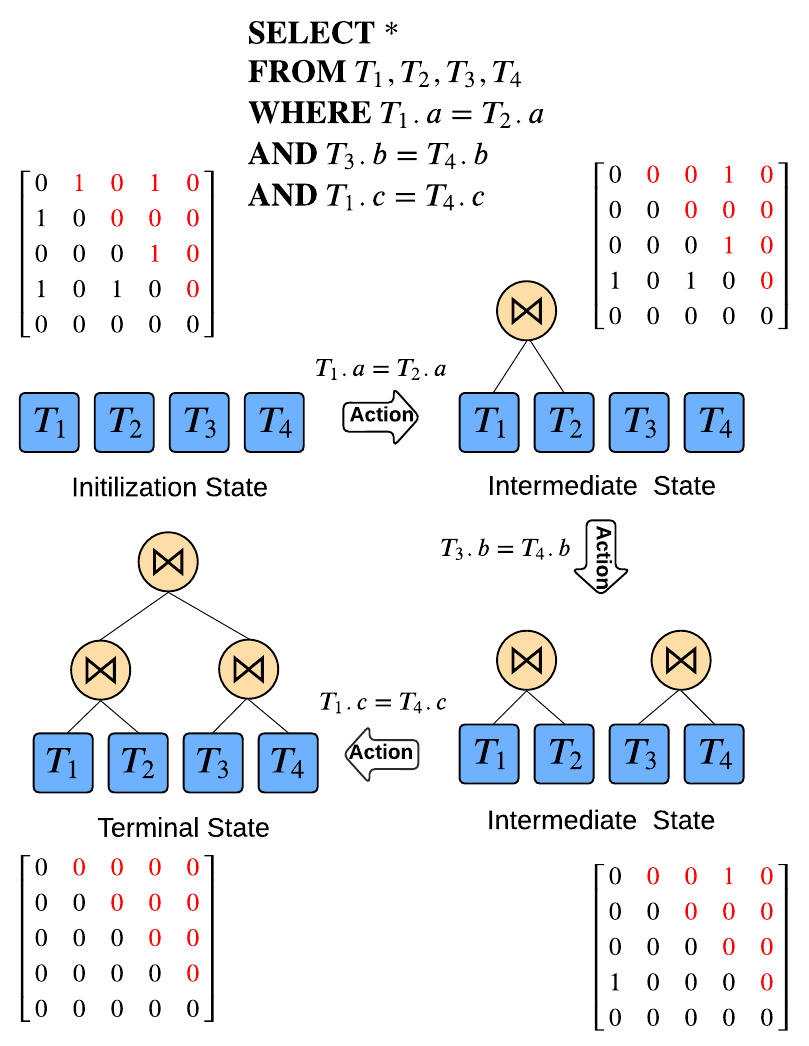}}
\vspace{-0.1in}
\caption{GTDD selects two tables joined together as an action until all tables are joined}
\label{fig_join_example}
\Description{Process of selecting the join order for a given query.}
\vspace{-0.1in}
\end{figure}

The join order selection problem is a critical task in query optimization within DBMS. It involves determining the most efficient sequence to execute joins between tables in a query, which significantly affects the query's execution time. Traditional approaches, such as dynamic programming and heuristic-based algorithms, struggle with scalability as the number of tables increases. Many researchers leverage DRL to address the limitations of traditional plan enumerators. The model utilizes the feedback to search for the optimal join orders and avoid suboptimal ones. The learned plan enumerator could incorporate the feedback as a guide to enable the DRL agent to learn through trial and error, thereby improving the quality of the generated join orders for queries. A DRL-based model contains several important components, and we can define them in the context of join order selection: 
\begin{itemize}
    \item \textbf{Agent:} The agent is the optimizer that interacts with the environment to learn the optimal sequence of join operations for executing a query.
    \item \textbf{Environment:} The environment provides the necessary context for the agent actions and returns feedback, which is the underlying DBMS.
    \item \textbf{State:} The state includes initial state, intermediate state and terminal state. Figure \ref{fig_join_example} demonstrates all the aforementioned states. At the initialization state, all tables are awaiting a join, and at each subsequent step, the agent selects an action to observe an intermediate state. Once all tables are joined, it observes the terminal state.
    \item \textbf{Action:} The action refers to choosing which two tables (or intermediate results) to join next.
    \item \textbf{Reward:} The agent does not observe the reward until reaching the terminal state. At the terminal state, the reward is calculated as an improvement ratio, comparing the baseline plan with the plan generated by the learned optimizer.
    \item \textbf{Memory Pool:} The memory pool is used to store the past status of the environment and the feedback from DBMS. At the end of the episode, the agent samples a mini-batch of experiences from the memory pool to update the neural network.
\end{itemize}

Although recent research has demonstrated remarkable results in applying DRL to the join order selection problem, DRL-based models still exhibit significant limitations:
    \begin{itemize}
        \item \textbf{Reward Design:} The reward function directly influences the agent's learning process and decision-making. An appropriately designed reward function will encourage the agent to prioritize actions that lead to lower costs, thus driving the agent toward optimal or near-optimal join orders.
        \item \textbf{Convergence Speed:} Faster convergence means the DRL model reaches a useful policy or solution more quickly. Quicker convergence allows the optimizer to be integrated into real-time query processing environments more effectively.
    \end{itemize}

\section{An Overview of GTDD}
 GTDD trains a learned optimizer using DRL with Tree-LSTM and GNN. The model takes a query as an input and outputs the desired join order for the input query. The DBMS then executes the query with a given join order and observes feedback(e.g. estimated cost), it is then used by GTDD as new training samples to improve itself. Similar to other research \cite{ref_DQ},\cite{ref_ReJoin},\cite{ref_RTOS} and \cite{ref_Jogger}, our work focuses on Select-Project-Join (SPJ) queries.
 Figure \ref{fig_join_example} demonstrates an entire process of join order selection for a query. At the initialization state, all tables are waiting for a join. At each subsequent step, the agent selects an action that is expected to be optimal, based on the largest estimated long-term reward. Once the action has been taken, the agent observes a different state. This process repeats until all tables are joined, which indicates that the execution plan is ready to be passed to the DBMS for execution. Once the DBMS provides feedback regarding the derived query plan, GTDD utilizes the feedback as a guide to update the agent.
 
\begin{figure}[t]
\centerline{\includegraphics[scale=0.5]{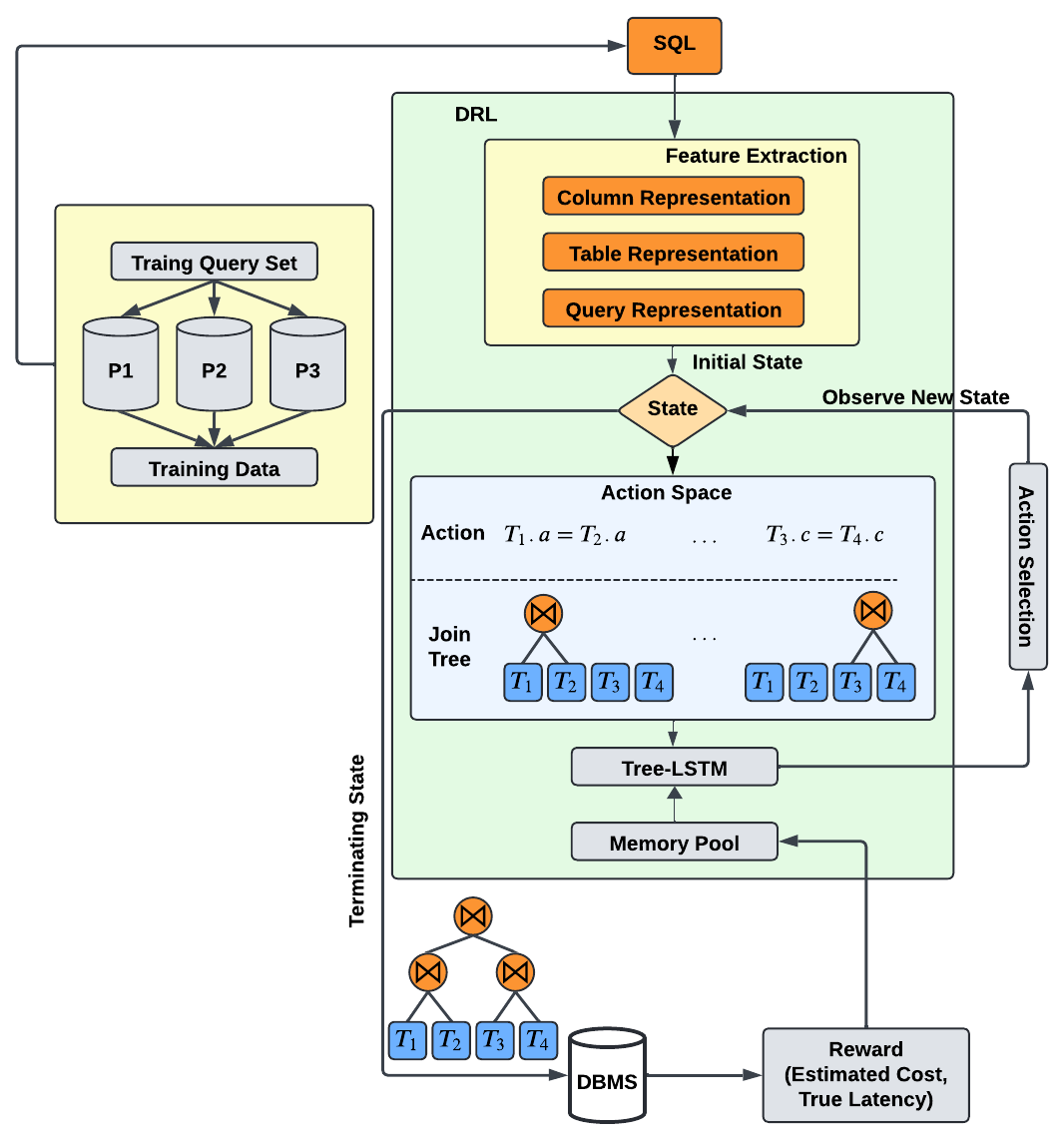}}
\vspace{-0.1in}
\caption{The GTDD framework}
\label{fig_framework}
\Description{The framefork of GTDD.}
\vspace{-0.1in}
\end{figure}

Figure \ref{fig_framework} demonstrates the framework of GTDD. The training query set is split into three partitions, and the training data is sampled from the partitions. The sampled query passes into the feature extraction component to learn three levels of representations: column representation, table representation, and query representation. These representations are then fed into different neural networks to acquire the initial state. The agent selects an action based on the state information, and the state after taking the selected action passes into the Tree-LSTM network to obtain a new intermediate state. This process repeats until the terminal state is reached, where all the tables are joined together. The join order is then passed to the underlying DBMS to obtain the reward (e.g. cost), and the result with the join order passes into the memory pool to update the model. GTDD is based on DRL, which aims to identify the optimal join order that produces the minimum cost by interacting with the environment. The framework is not only based on DRL but also on representation learning~\cite{ref_RepresentationLearning}. As shown in Figure \ref{fig_framework}, GTDD first extracts information and feeds it into different levels of representation learning. The learned representations are then fed into a DRL neural network, enabling the agent to select feasible actions under the given state.

\section{Representation Learning}
Capturing informative representations of input queries is crucial for enhancing the training process and determining the optimal join order. Representation Learning~\cite{ref_RepresentationLearning} encompasses a set of techniques aimed at learning features from raw data, transforming it into forms that can be efficiently utilized by machine learning algorithms. In our approach, we implement four distinct levels of representation learning, column-level, table-level, query-level, and state-level.

\subsection{Column Representation Learning}
 For each query $Q$, there are two types of predicates, join predicate and selection predicate. The join predicate refers to two tables that join together on the given column, and the selection predicate serves as a filter, eliminating data that do not satisfy the condition. Figure \ref{fig.column_representation} illustrates that each column $c$ in a table is represented by a vector of size 6. The elements of this vector are defined as follows:\\
    \centerline{$F(c) =[c_{\bowtie},c_{=},c_{<},c_{>},c_{\leq},c_{\geq}]$}
    where $c_{\bowtie}$=1 indicates that the column participates in a join, and the other operators refer to the selectivity of the column with respect to different operators. We treat the operator "BETWEEN" with both $c_{<=}$ and $c_{>=}$ for the upper and lower bounds, respectively. Additionally, a matrix $M(c)$ is defined with shape $[6, hs]$ for each column, where $hs$ is the size of the hidden layer and it is a learnable parameter that contains the information. The final column representation is $R(C) = F(C)*M(C)$.\\ 
    
    \textbf{Example 1:} Consider Query \ref{eq:query_example}, F($T_1.a$) = (0,0,0.4,0.6,0,0) since column $T_1.a$ has no join condition, "$=$", "$\leq$", nor "$\geq$" predicates. Consequently, we set $c_{\bowtie}$, $c_{=}$, $c_{\leq}$ and $c_{\geq}$ to 0. Assume column $T_1.a$ contains 100 unique data, ranging from 1 to 100. In this case, we set $c_{<}$ to $\frac{40}{100} = 0.4$ for $T_1.a < 40$ and $c_{>}$ to $\frac{60}{100} = 0.6$ for $T_1.a > 60$. The final column representation for $T_1.a$ is $F(T_1.a)*M(T_1.a)$. 
    \begin{equation}
\begin{split}
& \textbf{SELECT}\ T_1.c\\
& \textbf{FROM}\ T_1,T_2,T_3,T_4\\
& \textbf{WHERE}\ T_1.a < 40\\
& \hspace{0.1in}\textbf{AND}\ T_1.a > 60\\
& \hspace{0.1in}\textbf{AND}\ T_1.d\ \textbf{BETWEEN}\ 10\ \textbf{AND}\ 20\\
& \hspace{0.1in}\textbf{AND}\ T_2.b = T_3.b\\
& \hspace{0.1in}\textbf{AND}\ T_1.c = T_4.c; \label{eq:query_example}
\end{split}
\end{equation}

Our column representation captures selectivity information, which plays a crucial role in query optimization, as it estimates the number of rows returned by a join operator or a selection predicate. These estimates directly impact the efficiency of the query execution plan.

\subsection{Table Representation Learning}
    For each table $t$ with $n$ columns, the combination of column representations is used to represent the table. While extracting the information from the individual columns of the table is crucial, the correlations between the tables also provide valuable semantic information. Columns with indexed attributes are more likely to be involved in join operations. We leverage the schema of the database to extract correlation between tables and combine the information from both column representations and information of tables from schema to obtain the final representation of tables. 
   
    \begin{figure}[t] \centerline{\includegraphics[scale=0.15]{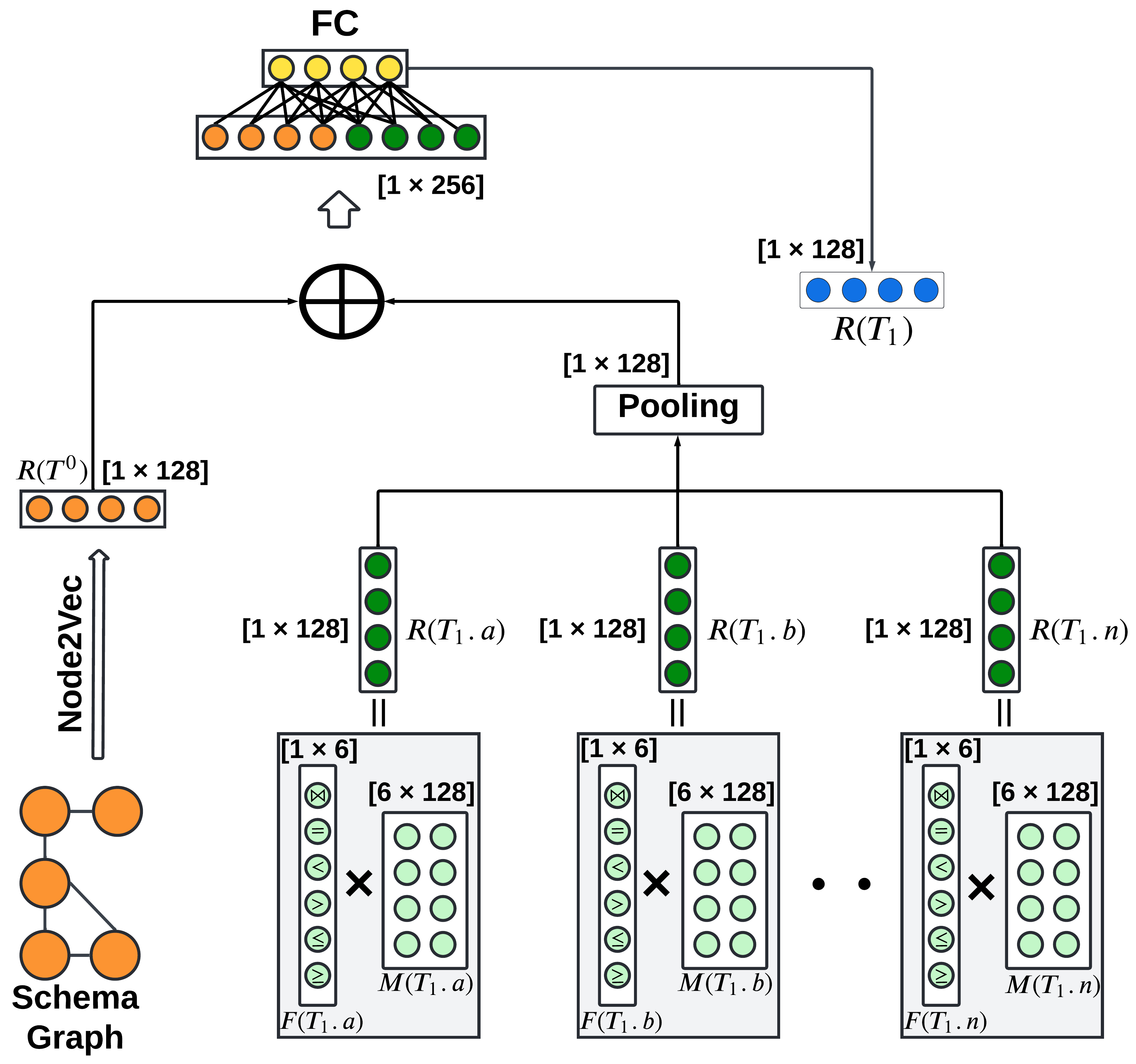}}
    \caption{Column representation and table representation}
    \label{fig.column_representation}
    \Description{Column representation and table representation of GTDD}
    \end{figure}
    
    However, the combination of column representations only captures the intrinsic information of the columns, while overlooking the contextual information provided by the schema. Schema contains relationships between tables, and we design an algorithm that converts the schema to a schema graph and captures additional embedding for each of the tables. The most common graph sampling strategies are breadth-first search (BFS) and depth-first search (DFS). BFS explores the graph level by level, starting from an unvisited node $v$ in the graph and moving outward. It visits all the neighbors of a node before moving to the next level of nodes. DFS explores as far along a branch as possible before backtracking. It goes deep into the graph, visiting a node's neighbors before backtracking and exploring other branches. These two methods are usually too extreme. Thus, we adapt Node2Vec~\cite{ref_node2vec}, which is a method for converting nodes on a graph into a vector embedding without any labels. Node2Vec is a biased random walk that uses transition probabilities to guide the walk, adjusting the likelihood of moving more like a BFS or DFS. We treat the database schema as an undirected graph $G = {(V^T, E^T)}$ where $V^T$ is a set of vertices (tables), and $E^T$ is a set of edges (primary-foreign key relations). By capturing the schema information, more insights can be gained about the overall schema graph. To be able to balance both strategies, node2Vec provides a biased random walk that uses two hyper-parameters $p$ and $q$ to control the walk, $p$ is the return parameter which controls the likelihood of returning to the previous node, higher $p$ denotes that the walk is less likely to return to the previous node, while $q$ is in-out parameter which determines the likelihood of visiting nodes that are further away versus closer neighbors, which is more likely a DFS-like exploration. As shown in Algorithm \ref{alg:node2vec}, we first construct a schema graph based on the primary-foreign key relationships, that randomly selects a node $T_i$ as the starting point and performs a walk in the graph. It selects the neighbor to move to on the next step based on probabilities 
    controlled by $p$ and $q$. 
    This process repeats until it reaches the maximum walk length. After many trails, the sequence of the walks is captured as elements for each table. Second, the algorithm learns the representation from the sequence collected in the first step. We treat the node in the sequence as a word, and the sequence as a sentence. We can take advantage of the well-learned language model \textit{Skip-Gram}~\cite{ref_skipgram} to learn the node representation. By applying these two methods, we have the schema embedding $R(T^0)$ as shown in Figure \ref{fig.column_representation}. 
    
    \begin{algorithm}[t]
    \caption{Schema Representation}\label{alg:node2vec}
    \begin{algorithmic}
    \Input
    \Desc{Database Schema}
    \Desc{maximum length of the walk $W$}
    \Desc{hyper-parameters $p$ and $q$}
    \EndInput
    \Output
    \Desc{Schema representations}
    \EndOutput
    \end{algorithmic}
    \begin{algorithmic}[1]
    \State Build a schema graph G = ($V^T,E^T$) based on the primary-foreign key relationships;
    \While{not reach the number of walks N}
    \For{node $n$ in $V^T$}
    \While{not reach the length of the walk W}
    \State select next node $Next$ based on:
    \State $Next$ = 
    $\begin{cases}
        \frac{1}{p} \ \texttt{move to previous node}\\
        \frac{1}{q} \ \texttt{move to neighbour node}
    \end{cases}$
    \EndWhile
    \EndFor
    \EndWhile
    \State Perform the Skip-gram algorithm and use an average pooling layer to get the correlation embedding $R(T^0)$ for each table in the database
    \end{algorithmic}
    \end{algorithm}
    
    The column representation continues to provide indispensable information, as it is the fundamental aspect of table representation. To ensure the final table representation contains information from both schema and the corresponding column, we employ a pooling layer that extracts information from column embedding and concatenates them together:
    
    \centerline{$R(T) = FC(Pooling(R(T_{c1}),...,R(T_{cn})) \bigoplus R(T^{0})$)}
    where $R(T^0)$ is the feature of table from schema embedding and $R(T_{ci})$ is the i-th column embedding for table $T$. To be able to maintain the same size for further computation, we deploy a fully connected layer as Figure \ref{fig.column_representation} demonstrates.
    
\subsection{State Representation Learning}
The state is a crucial component in DRL which affects the learning performance. An informative state representation could significantly enhance the performance of the model. The state representation is the concatenation of the query presentation $R(q)$, learned by GNN, and the current join forest $R(\mathcal{F})$, learned by Tree-LSTM. To be consistent, the state representation $R(s)$ is the join state, where $R(s) = R(q) \oplus R(\mathcal{F})$. 
    \subsubsection{Query Representation}
\begin{figure}[t]
    \centerline{\includegraphics[scale=0.22]{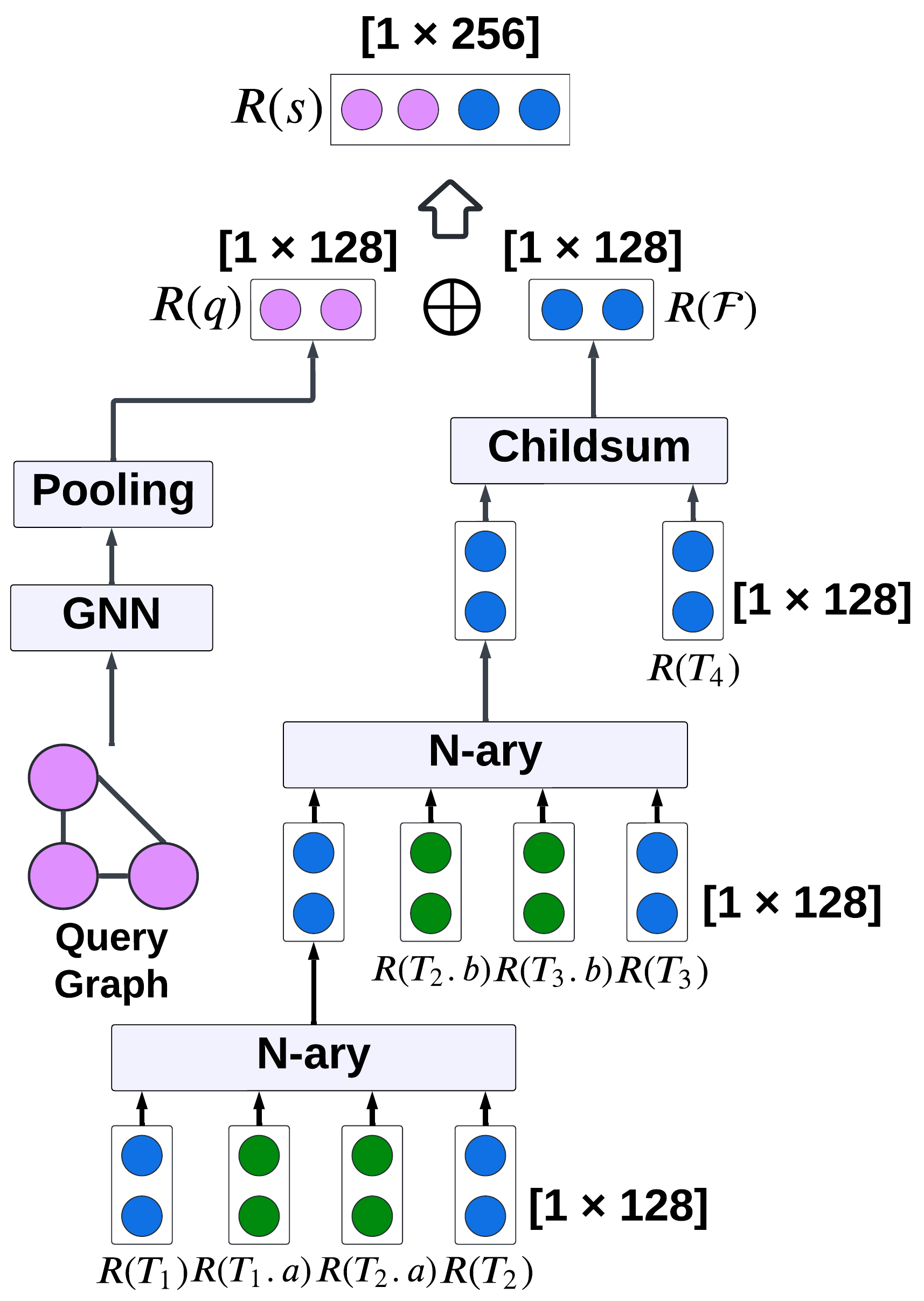}}
    \caption{State representation}
    \label{fig.state_representation}
    \Description{State representation of GTDD}
    \end{figure}

    GTDD uses a join matrix $m$ with size $n\times n$ as an adjacency matrix to create an undirected join graph to learn the query representation, where $n$ represents the number of tables in the database and $m_{i,j}$ denotes whether tables $T_i$ and $T_j$ participate in a join in the given query, with a value of 1 indicating join exists and 0 otherwise. By doing so, GTDD not only captures the link information between the tables but also captures the intrinsic table representation, as Figure \ref{fig.state_representation} shows. To be more specific, we employ two-layer transformerconv~\cite{ref_transformerconv}, which is a self-attention mechanism that dynamically weighs the influence of neighboring nodes in a graph, we then use global pooling to obtain the query graph representation.
    \subsubsection{Join Tree, Join Forest and the State Representation}
    To capture the hierarchical structure inherent in tree-structured data and the data features from a long sequence, we implement the Tree-LSTM~\cite{ref_treelstm} model. The model captures compositional semantics by recursively combining representations of child nodes to compute the representation of their parent node, thereby enabling the model to capture informative representations and the sequence of the join order. We use two distinct types of Tree-LSTM, the child-sum Tree-LSTM and the n-ary Tree-LSTM.\\
    \textbf{Child-sum Tree-LSTM} can be considered as a pooling layer that does not consider the sequence of its children. It selects representative information from its children and obtains the overall representation. For a given node $j$ with multiple children $\alpha_{j,k}$, the model computes the sum of representations from all child nodes to construct its own representation.\\
    \textbf{N-ary Tree-LSTM} is different than child-sum Tree-LSTM, it has a fixed number of children and considers the order of its children, which could capture the sequence of the join order. For each internal node $j$, there are exactly $N$ children $\alpha_{j,k}$. The representation of each child $\alpha_{j,k}$ is computed separately and each child node has its own weight matrix. 
    \paragraph{Join Tree Representation}
    In order to obtain the representation of the forest $F$ which is composed of several join trees {$\mathcal{T}_1,\mathcal{T}_2,...,\mathcal{T}_n$}, we employ n-ary Tree-LSTM since each tree should have the fixed number of children. The join tree is comprised of two distinct types of nodes: leaf nodes and internal nodes. A leaf node represents a table or a column, while an internal node corresponds to a join that is composed of 4 nodes, namely, $\alpha_0$,$\beta_0$,$\beta_1$,$\alpha_1$. The nodes $\alpha_0$ and $\alpha_1$ represent tables or join trees, while the nodes $\beta_0$ and $\beta_1$ represent the corresponding columns in the join.
    
    \begin{algorithm}[t]
    \caption{Encodetree (Node n)}\label{alg:encodetree}
    \begin{algorithmic}
    \Input
    \Desc{Node n}
    \EndInput
    \Output
    \Desc{Representation of Join tree}
    \EndOutput
    \end{algorithmic}
    \begin{algorithmic}[1]
    \If{n is a leaf Node}
        \State h = R(n)
        \State m = Zeros\textunderscore init()
        \State \Return h,m
    \Else
        \State $h_{\alpha_0},m_{\alpha_0}$ = Encodetree($\alpha_0$)
        \State $h_{\alpha_1},m_{\alpha_1}$ = Encodetree($\alpha_1$)
        \State $h_{\beta_0},m_{\beta_0}$ = Encodetree($\beta_0$)
        \State $h_{\beta_1},m_{\beta_1}$ = Encodetree($\beta_1$)
        \State \Return n-aryUnit($h_{\alpha_0},m_{\alpha_0},h_{\alpha_1},m_{\alpha_1},h_{\beta_1},m_{\beta_1}$)
    \EndIf
    \end{algorithmic}
    \end{algorithm}
    
    As illustrated in Figure \ref{fig.state_representation}, when two tables are joined together, we use the n-ary Tree-LSTM to construct the join tree. The leaf node is either a column or a table, while the internal node has 4 children, denoted as $\alpha_0$,$\beta_0$,$\beta_1$,$\alpha_1$. The nodes $\alpha_0$ and $\alpha_1$ represent two join trees (tables) in a join predicate, while the nodes $\beta_0$ and $\beta_1$ represent two columns that participate in the join. As illustrated in Algorithm \ref{alg:encodetree}, the tree is constructed recursively. Similar to the traditional LSTM, the Tree-LSTM comprises two key components: $h_j$ and $c_j$, which represent the hidden state and cell state, respectively, for each tree node $j$. The hidden state acts as a short-term memory and captures recent information that is immediately relevant to the current prediction. The cell state acts as a long-term memory and provides the network with the ability to remember information for long durations. 
    \begin{itemize}
        \item If node $j$ is a leaf node, $h_j = R(j)$, $j$ could be either a column or table
        \item If node $j$ is an internal node representing a join, it must have 4 child nodes $\alpha_0$, $\beta_0$, $\beta_1$, $\alpha_1$.The representation of these 4 children will feed into the n-ary unit in n-ary Tree-LSTM to get the representation of node $j$
    \end{itemize}
    
    We use hidden state $h_j$ to represent the intermediate output of the n-ary Tree-LSTM, which also represents a join tree $\mathcal{T}$ such that $R(\mathcal{T})=h_j$.
    \paragraph{Join Forest Representation} The join forest is composed of a set of trees, where join forest $\mathcal{F} = \{\mathcal{T}_1,\mathcal{T}_2,...,\mathcal{T}_n\}$. since the number of tables involved in a query can vary, the number of trees also varies. An n-ary Tree-LSTM would not be suitable in this case, so we apply child-sum Tree-LSTM to handle the issue as it disregards the number of children. Similarly to the n-ary Tree-LSTM, the child-sum Tree-LSTM also has join states $h_j$ and $c_j$. However, it differs by summing all the hidden states $h_j$ for each tree in the forest to obtain a new hidden state $h^{root}$, which serves as the intermediate output of the child-sum Tree-LSTM. The representation of a join forest $\mathcal{F}$ is thus given by $R(\mathcal{F})=h^{root}$.
    \paragraph{State Representation} As Figure \ref{fig.state_representation} shows, the state representation $R(s)$ would be the concatenation of $R(q)$ and $R(\mathcal{F})$ such that $R(s) = R(q)\bigoplus R(\mathcal{F})$ and feed into the Dueling-DQN as state information. 
\section{Deep Reinforcement Learning}
The join order selection problem can be easily mapped into RL as denoted in Section IV, and we have described the presentation for state representation. The next goal is to discuss how RL works in join order selection.
\subsection{DQN and Dueling-DQN}

\begin{figure*}[t]
    \centering
    \begin{subfigure}[b]{0.55\textwidth}
        \centering
     \includegraphics[scale=0.2]{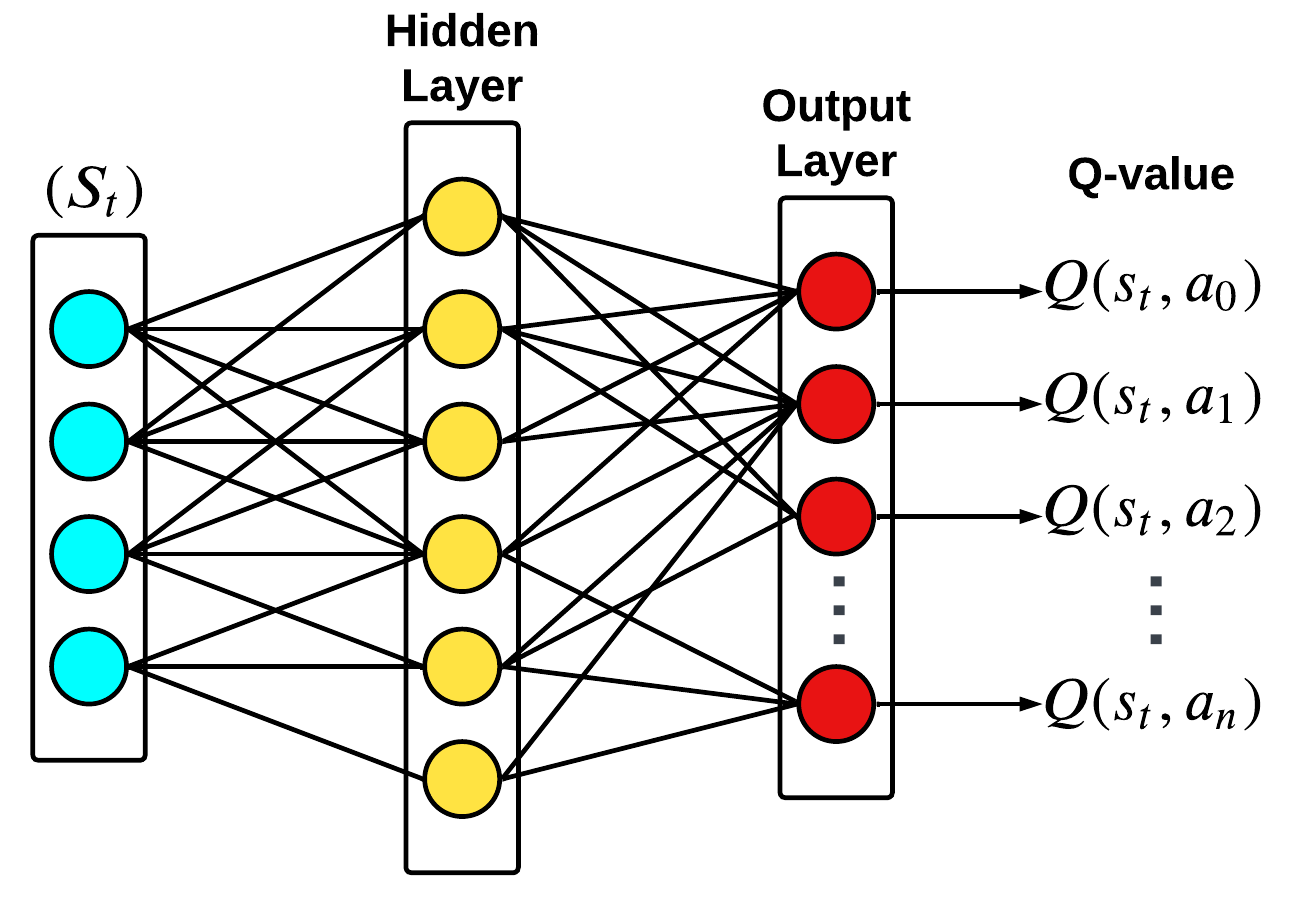}
        \caption{Deep Q Network}
        \label{fig.dqn}
    \end{subfigure}
    \hspace{-0.8in} 
    \begin{subfigure}[b]{0.55\textwidth}
        \centering
    \includegraphics[scale=0.2]{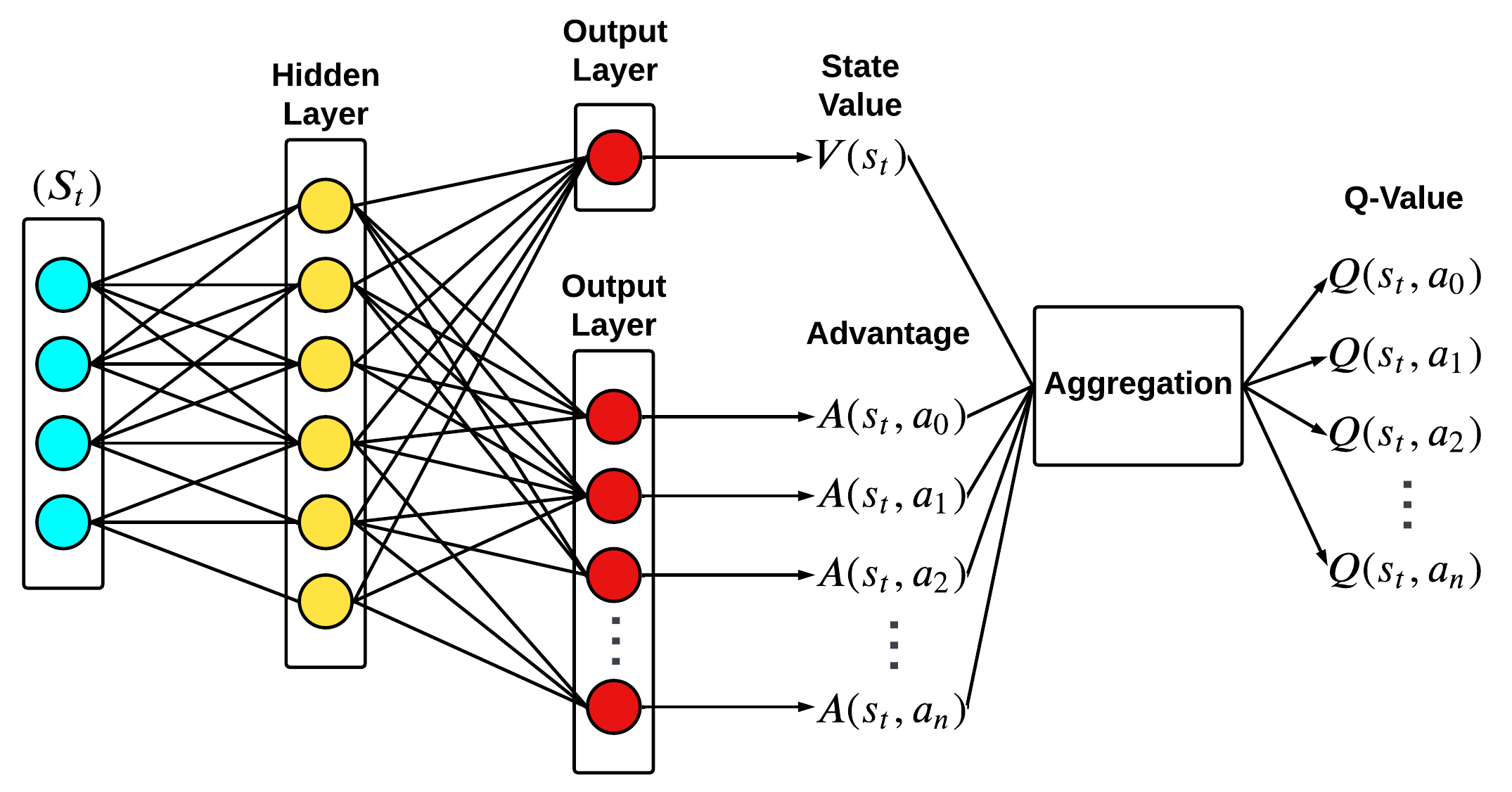}
        \caption{Dueling Deep Q Network}
        \label{fig.dueling_dqn}
    \end{subfigure}
    \caption{DQN vs Dueling-DQN}
    \Description{Difference between vanilla DQN and Dueling-DQN}
\end{figure*}

The majority of work like DQ~\cite{ref_DQ}, RTOS~\cite{ref_RTOS} and JOGGER~\cite{ref_Jogger} adopt the classical reinforcement learning strategy that fits the best in discrete action space, DQN~\cite{ref_DQN}. Although their work demonstrates remarkable results, the training stability and convergence time remain potential drawbacks, and none of these works use different RL methods to stabilize the training process. We discover one potential solution to stabilize the training process, Dueling-DQN~\cite{ref_DuelingDQN}, which is an enhancement of the DQN architecture. It separates the estimation of the state value and the advantage of each action, which can lead to better learning and decision-making. As dueling-DQN is a variant of DQN, we provide an explanation of DQN and then explain how dueling-DQN differs from DQN. As Figure \ref{fig.dqn} demonstrates, DQN is a single neural network that directly estimates the q-values $Q(s,a)$ for each action in a given state, and the q-value is used to guide the agent to select best action under the given state. Different than DQN, dueling-DQN is a modified neural network that separates the estimation of the state value and the advantage of each action. As Figure \ref{fig.dueling_dqn} demonstrates, dueling-DQN is divided into two streams, value stream and advantage stream. 
\begin{equation}
    Q(s,a;w) = V(s;\theta)+(A(s,a;\theta)-\frac{1}{|\mathcal{A}|}\sum_{a'\in \mathcal{A}}A(s,a';\theta))
\label{eq:dueling_dqn}
\end{equation}
Equation \ref{eq:dueling_dqn} demonstrates the process of aggregation of value stream and advantage stream, value stream, estimates the value function $V(s)$, which represents the overall value of being in state $s$ regardless of the action taken. Advantage stream estimates the advantage function $A(s,a)$, which represents the relative benefit of taking action a compared to other actions in state $s$, and $\frac{1}{|\mathcal{A}|}\sum_{a'\in \mathcal{A}}A(s,a';\theta)$ is the normalization term that computes the average advantage over all possible actions $a'$ in state $s$. This term ensures that the advantage function is normalized, meaning the advantages are centered around zero, which helps to stabilize the learning process. Parameter $\theta$ represents the parameters of the shared network between the two streams, and $\mathcal{A}$ is the set of all possible actions. 
Both of the networks take the state $s$ as input and eventually output q-values for all possible actions, where the action space is defined as $\mathcal{A}_s=\{a_1,a_2,...,a_n\}$. In order to achieve a balance between exploitation and exploration, the action is controlled by epsilon $\epsilon$, with the agent randomly selecting an action with the probability $\epsilon$ like:
\begin{center}
    $a = $ 
    $\begin{cases}
      Random\ Action & with\ probability\ \epsilon\\
      argmax_aQ(s,a\in\mathcal{A}) & with\ probability\ 1-\epsilon
    \end{cases}$
\end{center}

Both DQN and dueling-DQN have two networks, the prediction network $Q^{predict}$ and the target network $Q^{target}$. $Q^{predict}$ approximates the q-values for each state-action pair. This network is used to select actions during the exploration of the environment. $Q^{target}$ is a copy of $Q^{predict}$, but its parameters are updated less frequently. This separation helps to stabilize the training process by providing consistent targets for the updates of $Q^{predict}$. Both networks are initialized with the same structure and the same parameters. The target network periodically copies the parameters of the prediction network as an update. The loss function is the standard MSE, which aims to minimize the gap between $Q^{predict}$ and $Q^{target}$:

\begin{equation}
    L(\theta) = (y-Q^{predict}(s,a;w))^2
\label{eq:loss}
\end{equation}

\begin{equation}
    y = r+\gamma max_{a'}Q^{target}(s',a';w)
\label{eq:loss_y}
\end{equation}

Equation \ref{eq:loss} is the loss function that shows the difference between the predicted value and the target value.
The primary goal is to minimize the difference between the predicted value and the target value.
Equation \ref{eq:loss_y} specifies the target value
calculated as the sum of the reward and the discounted maximum q-value of the next state, where a separate target network is employed to enhance the stability of learning. The parameter $\gamma$ represents the discount factor, which controls the weighting of future rewards relative to immediate rewards; a value close to 0 implies that the agent favors immediate rewards over those in the future. Through iterative interactions with the environment, the agent incrementally learns an optimal join order.

\subsection{Reward}

    \begin{figure}[t]
    \centerline{\includegraphics[scale=0.35]{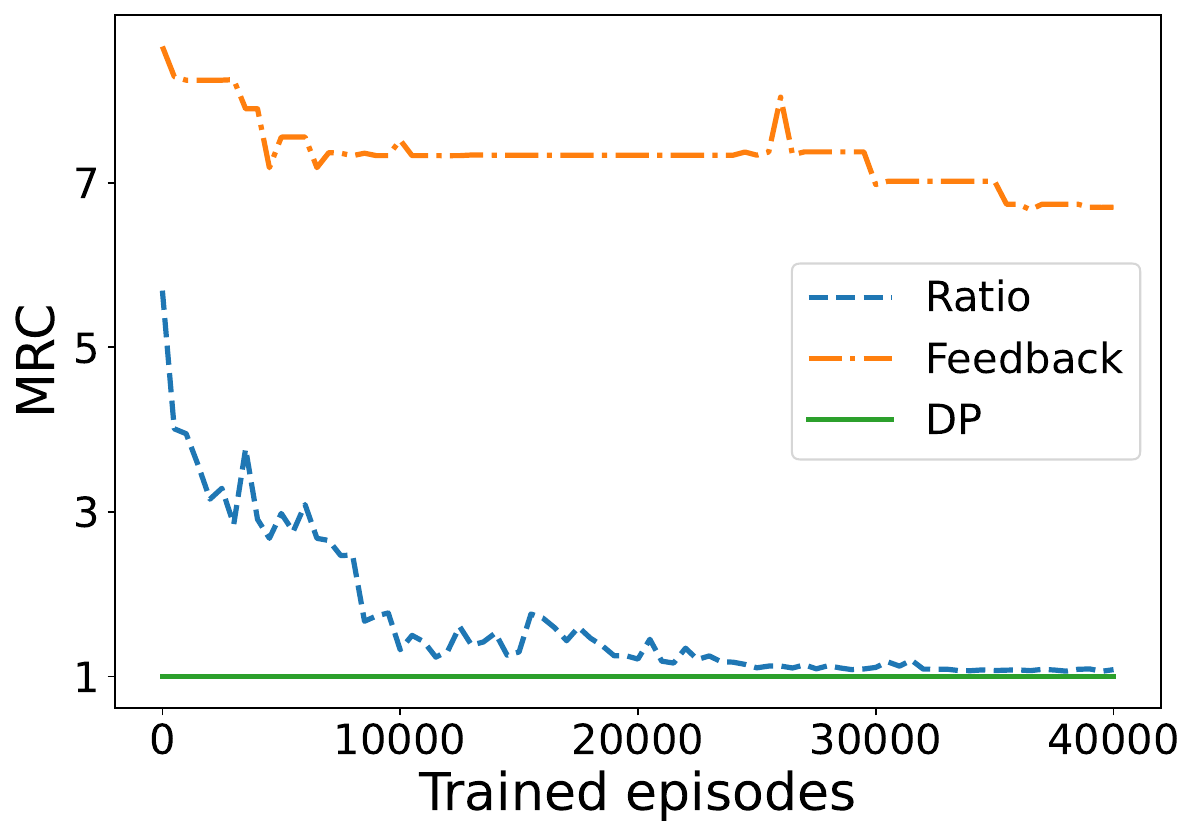}}
    \vspace{-0.1in}
    \caption{Feedback as reward vs ratio as reward}
    \label{fig.old_reward_vs_new_reward}
    \Description{different reward function implementation}
    \vspace{-0.1in}
    \end{figure}
    
The reward function is a crucial component in RL as it directly impacts the behavior and performance of the agent. It defines the agent's objective by providing feedback on its actions and directing it towards the desired outcomes. Recent research, including DQ~\cite{ref_DQ}, ReJOIN~\cite{ref_ReJoin}, and RTOS~\cite{ref_RTOS}, establishes a reward of 0 for all intermediate states. For the terminal state, the reward value is set to the feedback (e.g. cost or latency) of the entire execution plan as provided by the DBMS. However, as Figure~\ref{fig.old_reward_vs_new_reward} shows, using the feedback as reward leads to selecting suboptimal query plans.
We infer that this may be due to the varying ranges of feedback for different queries. Simple queries tend to have significantly lower cost and latency compared to complex queries, leading to misestimation of the q-value in states. Similar to the reward setting we aforementioned, we set the reward to 0 for all the intermediate states, and for the terminal reward, we adopt a different approach. 
We express the reward as the logarithm a ratio of the feedback of the plan selected by DP to the feedback of the plan selected by RL, as shown in Equation \ref{eq:loss_reward}:

\begin{equation}
    r = log(\frac{DP_{feedback}}{GTDD_{feedback}})
\label{eq:loss_reward}
\end{equation}



This formula is used to compare the relative costs of two plans, where we use the plan selected by DP as the baseline to calculate the improvement ratio as the reward function.
%
The logarithm provides meaningful information regardless of the absolute magnitudes of the feedback (e.g. cost). It helps the agent to efficiently discern the relative merits of different actions or states, especially in environments with widely varying costs.
The feedback of the plans selected by DP and GTDD are denoted by $DP_{feedback}$ and $GTDD_{feedback}$ respectively. The estimated cost serves as feedback in the cost training phase, while latency serves as feedback in the latency tuning phase.\\


\textbf{Example 2:} For the sake of illustration, let us assume that $GTDD_{feedback}$ = 100 and $DP_{feedback} = 90$. In this case, the reward $r(a)$ is given by the expression $log(\frac{90}{100}) = -0.046$, which is a negative value. This is because the execution plan of DP has a superior performance compared to GTDD. If we assume that $GTDD_{feedback} = 90$ and $DP_{feedback} = 100$, then the reward $r(a)$ is equal to $log(\frac{100}{90}) = 0.046$ which is a positive reward since GTDD obtain better performance. This reward function encourages the agent to find a better execution plan that results in positive rewards. 

\subsection{Action Mask}

The output of DRL is the q-value for all possible actions, which is usually not practical to obtain in real-life tasks. However, this overhead can be reduced as some of actions are invalid under certain scenarios (e.g. joins that lead to Cartesian products).
In GTDD we employ an action mask mechanism~\cite{ref_actionmask}, which allows for the implementation of a mask that filters out invalid actions, with a value of either 0 or 1.
This mechanism prevents the agent from taking invalid (or highly expensive) actions, allowing it to focus its learning capacity on the valid ones. This in turn leads to faster convergence and inference.\\

\textbf{Example 3:}
Recall the example shown in Figure ~\ref{fig_join_example}, and imagine the database contains 5 table $\mathcal{T}=\{T_1,T_2,T_3,T_4,T_5\}$. The matrices shown at each step represent the action mask where all invalid actions are set to 0 and valid ones are set to 1. The size of the mask matrix $\mathcal{M}$ is $N \times N$ where N is the number of tables in the database since there are $N \times N$ possible actions. In the initialization state, the values in the first row correspond to the join predicates connecting table $T_1$ to the other tables. For instance, the 1 in the second column means table $T_1$ could be joined with $T_2$, and the 1 in the fourth column means table $T_1$ could be joined with $T_4$. Since in the join ordering phase, A $JOIN$ B is equivalent to B $JOIN$ A, the size of the action space is reduced to $\frac{N\times(N-1)}{2}$.


\section{Experimental Study}
In this section, we present the experimental study we designed to evaluate the performance of GTDD. The primary objective is to assess how GTDD performs compared to state-of-the-art techniques. We begin by introducing the datasets used, and the evaluation metrics employed and the baselines of comparison. We continue by presenting a series of experimental results that demonstrate the advantages of GTDD over state-of-the-art techniques. GTDD is fully implemented~\cite{ref_code}.
\subsection{Experiment Setup}
\subsubsection{Computing Environment}
All of the experiments are run on an Intel(R) Core(TM) i7-9700K CPU @ 3.60GHz, and NVIDIA GeForce RTX 3070 Ti. We deploy the experiment using PostgreSQL on Windows 10.
\subsubsection{Datasets}
The experiments are conducted on two publicly available datasets. The Join Order Benchmark (JOB)~\cite{ref_JOB} and TPC-H~\cite{ref_tpch}. For each of the datasets, we select 90\% of the queries for training and the rest for testing.
\paragraph{Join Order Benchmark (JOB)~\cite{ref_JOB}} JOB is a real-world dataset derived from the Internet Movie Database (IMDB), which is a comprehensive database of information related to films, television series, actors, etc. JOB was specifically designed to evaluate the performance of different join order optimization strategies. JOB contains 113 queries derived from 33 templates, with each query containing a varying number of joins, ranging from a minimum of 4 to a maximum of 21. The total size of the database is 3.6G. Moreover, the schema of JOB has 21 tables and 108 columns in total, and we have used the schema file to construct the schema graph to capture the relations between tables.
\paragraph{TPC-H~\cite{ref_tpch}} The TPC-H dataset is a standard benchmark dataset used to evaluate the performance of DBMS, specifically designed to simulate a decision support system environment. The total size of the database is 4G. Moreover, the schema of TPC-H has 8 tables and 61 columns in total. 
\subsubsection{Baseline} We compare GTDD with both the traditional methods and DRL-based methods.
\paragraph{RTOS} RTOS~\cite{ref_RTOS} is a DRL-based approach that constructs a join forest with Tree-LSTMs to capture the sequential join information. This is then implemented by DQN to train the agent.
\paragraph{JOGGER} JOGGER~\cite{ref_Jogger} introduces a lightweight, tailored-tree-based attention module designed to effectively capture join orders with fewer parameters to tune compared to Tree-LSTM~\cite{ref_treelstm}. Furthermore, it employs DQN to train the agent.
\paragraph{GTD} To evaluate the effectiveness of dueling-DQN, we replaced the dueling-DQN component with a standard DQN in the GTDD framework, naming the resulting model GTD. GTD has the exact same implementation as GTDD, with the only difference being in the reinforcement learning component.
\paragraph{Dynamic Programming (DP)} DP~\cite{ref_DP} is a built-in technique in PostgreSQL, which is designed to find the optimal plan with the lowest cost. This method is controlled by a parameter 'geqo\_threshold', which determines the extent to which the dynamic programming approach is activated. By setting this parameter to a value greater than the number of tables involved in a query, the dynamic programming approach is activated.
\subsubsection{Metrics} We adopt the same metrics used in previous work to evaluate the performance of GTDD. Mean Relative Cost (MRC) is used to evaluate the performance during the cost training phase, while Geometric Mean Relevant Latency (GMRL) is used to evaluate the performance during the latency tuning phase. 
\paragraph{Mean Relative Cost (MRC)}
We take DP as the baseline for calculating MRC. An MRC value of 1 indicates that the plan generated by GTDD has the same cost as the plan generated by DP, while a smaller value indicates a better plan compared to DP. Equation ~\ref{eq:MRC} demonstrates the calculation of MRC, where $Q$ represents the entire dataset, and $q$ is a query within the set. The estimated cost for query $q$ from GTDD is denoted by $cost(q)$, while the estimated cost for query $q$ from DP is represented by $cost_{DP}(q)$. $|Q|$ represents the size of the dataset.
\begin{equation}
    MRC = \frac{\sum_{q \in Q}\frac{cost(q)}{cost_{DP}(q)}}{|Q|}
\label{eq:MRC}
\end{equation}
\paragraph{Geometric Mean Relevant Latency (GMRL)}
GMRL is used to evaluate the latency, which reflects the real-world performance of the model, as minimizing true latency is the primary goal. We have also selected DP as the baseline. The reason for adopting a different metric is that, after tuning with latency, the model generates better plans than DP. MRC would not be suitable to capture the relative performance. GMRL tends to provide a more balanced measure that reduces the impact of extremely high or low latency values, which might skew the results using the mean value. Equation ~\ref{eq:GMRL} demonstrates the calculation GMRL.
\begin{equation}
    GMRL = (\prod_{q \in Q}\frac{Latency(q)}{Latency_{DP}(q)})^{\frac{1}{|Q|}}
    \label{eq:GMRL}
\end{equation}
\subsection{Experimental Results}
Directly using cost as the reward function did not perform well in our case. For the sake of fairness, all the related work employs the reward function we propose.
\subsubsection{k-fold cross validation}
We performed 11-fold cross-validation on the JOB dataset, as it contains 33 different templates. In each fold, we included 3 templates in the test set and the remaining 30 templates in the training set.

\begin{figure*}
    \centering
    \begin{subfigure}[b]{0.5\textwidth}
        \centering
        \includegraphics[scale=0.33]{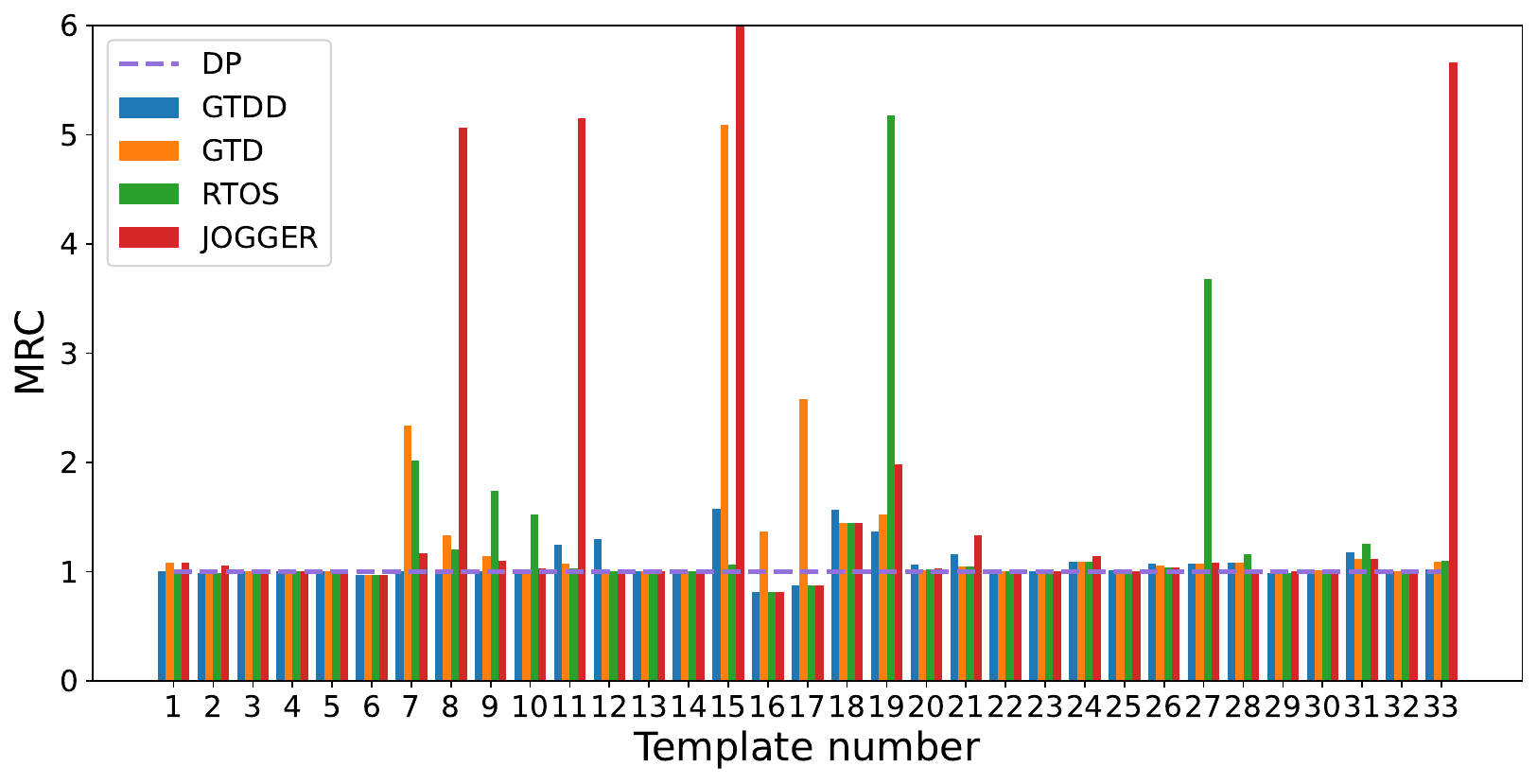}
        \caption{Per-template result using MRC, evaluated on the test set}
        \label{fig.10fold_job}
    \end{subfigure}%
    ~
    \begin{subfigure}[b]{0.5\textwidth}
        \centering
        \includegraphics[scale=0.33]{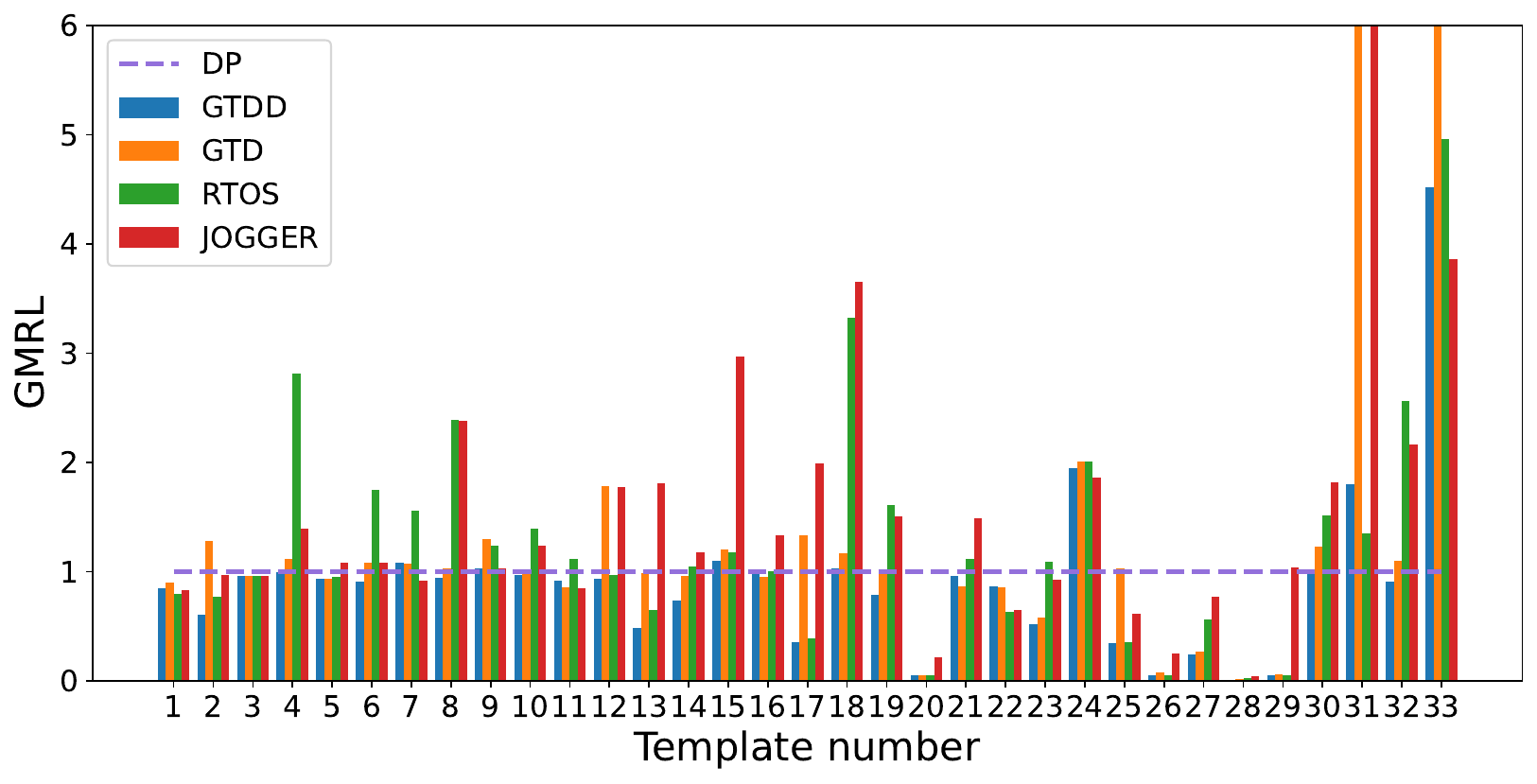}
        \caption{Per-template result using GMRL on JOB}
        \label{fig.latency_tuning_pattern}
    \end{subfigure}
    \caption{Per-template result in the cost training phase and latency tuning phase}
    \Description{Result in cost training phase and latency tuning phase based on templates for JOB}
\end{figure*}
Figure \ref{fig.10fold_job} presents the test set result in 11-fold cross-validation. All methods achieve satisfactory results on most templates. However, RTOS exhibits suboptimal performance on templates \#19 and \#27, while JOGGER performs inadequately on templates \#8, \#11, \#15, and \#33, which are among the most complex templates. Notably, GTDD demonstrates superior performance in terms of generalization compared to other approaches, even on these challenging templates. The main purpose of using k-fold cross-validation is to help ensure that the model generalizes well to unseen data. 
\subsubsection{Cost Training} We compare GTDD with the baseline methods discussed in the previous subsection in terms of MRC. During the cost training phase, we evaluate performance using MRC and compare the training times. Furthermore, we extract one entire template (template \#10) from JOB to test the performance of the model on unseen queries. In the latency tuning phase, we evaluate the performance using GMRL, and we also extract one entire template to test the performance of the model on unseen queries.
\begin{figure*}[htbp]
    \centering
    \begin{subfigure}[b]{0.32\textwidth}
        \includegraphics[scale=0.288]{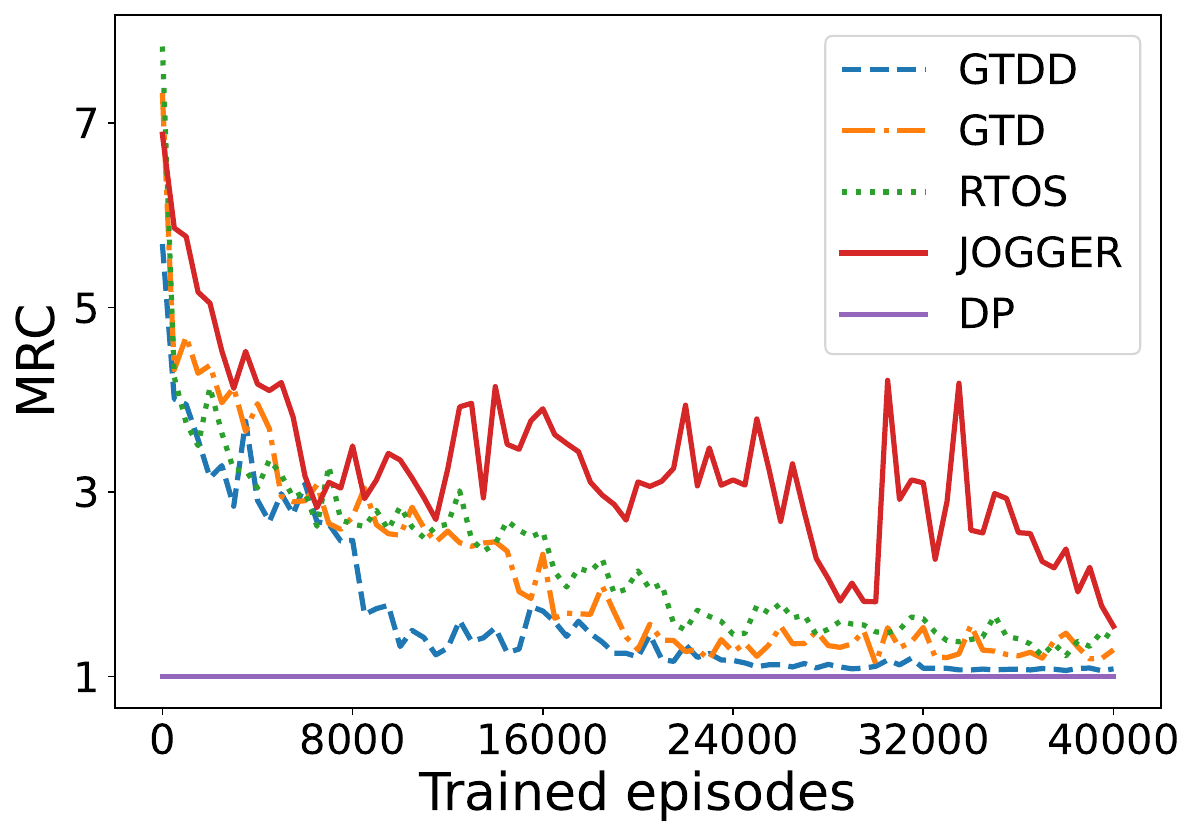}
        \caption{Cost training curve on JOB}
        \label{fig.cost_training}
    \end{subfigure}
    \hfill
    \begin{subfigure}[b]{0.32\textwidth}
        \includegraphics[scale=0.288]{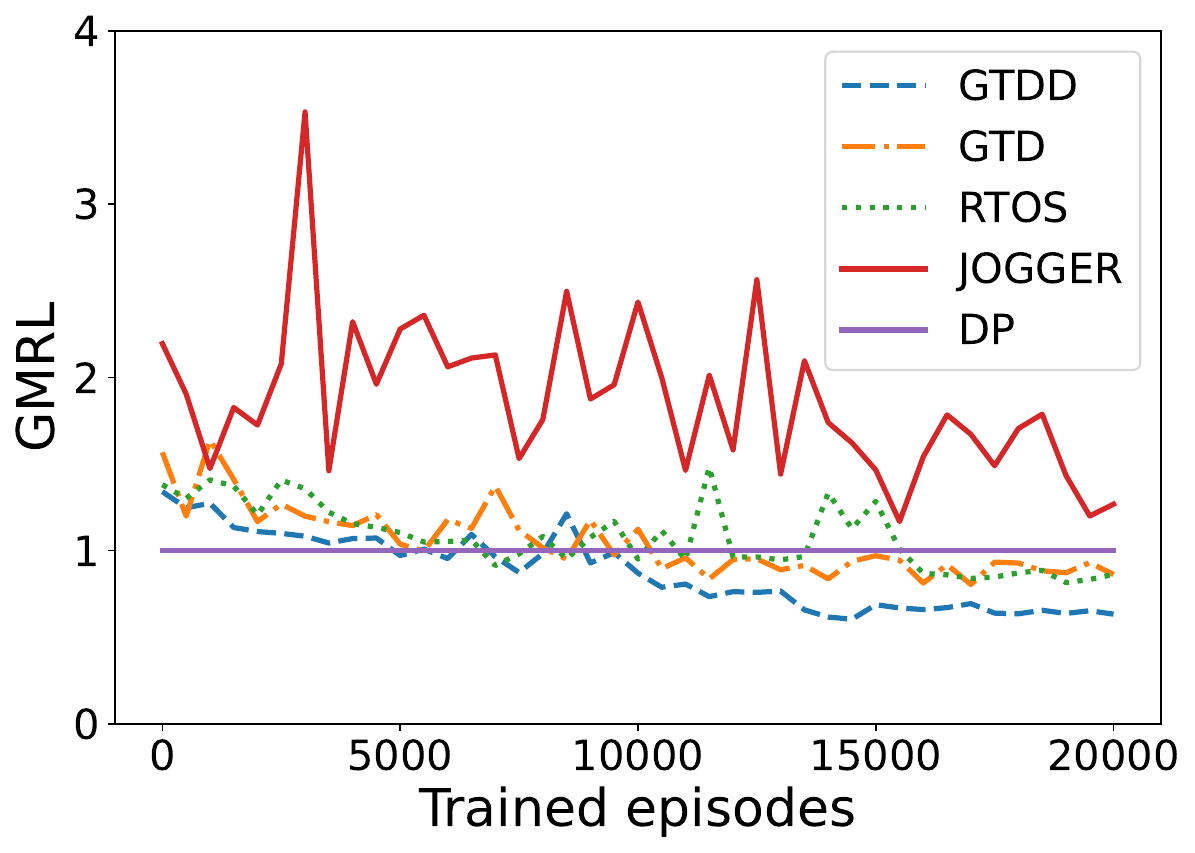}
        \caption{Latency tuning curve on JOB}
        \label{fig.latency_tuning}
    \end{subfigure}
    \hfill
        \begin{subfigure}[b]{0.32\textwidth}
        \includegraphics[scale=0.288]{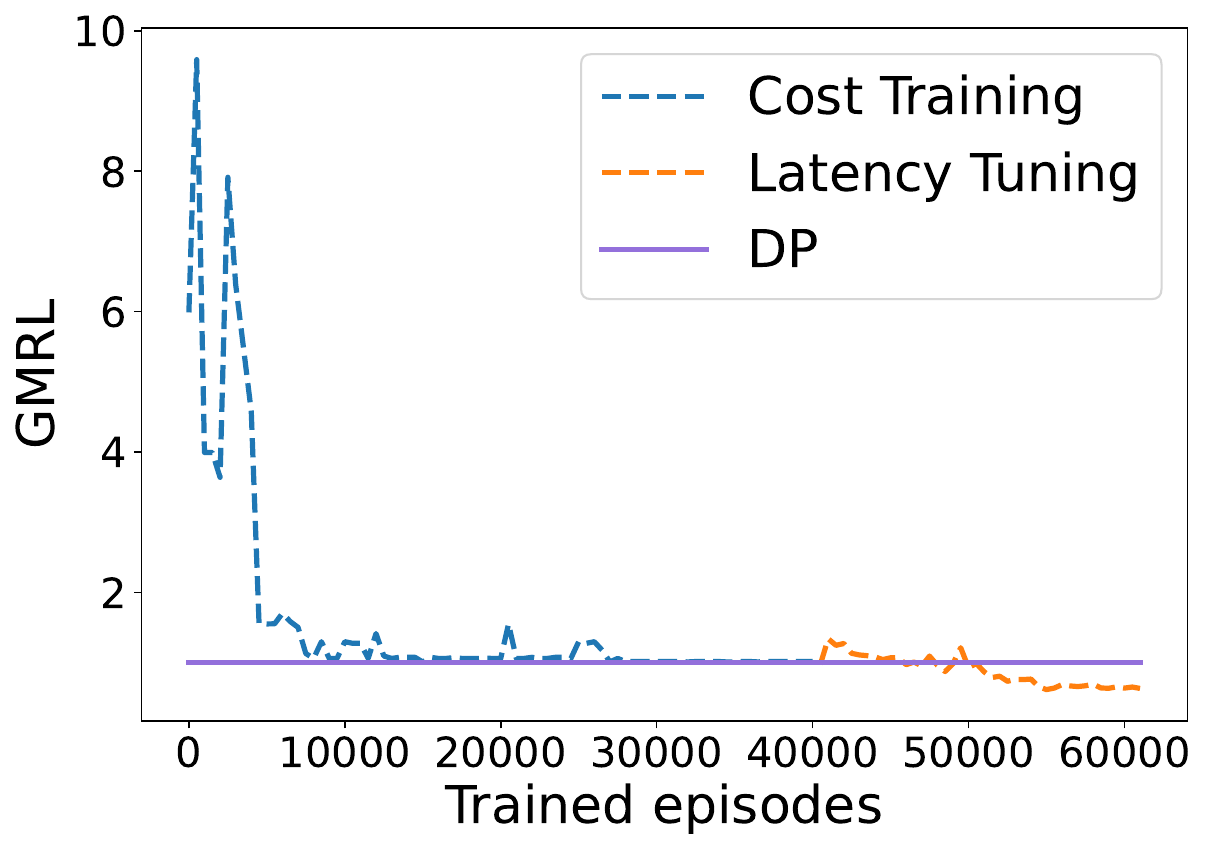}
        \caption{Cost + latency curve on JOB}
        \label{fig.training_tuning}
    \end{subfigure}
    \caption{Training curves in the cost training phase and latency tuning phase}
    \Description{Training curves of GTDD in both cost training phase and latency tuning phase}
\end{figure*}
Figure \ref{fig.cost_training} depicts the training curve on JOB dataset. Both DRL-based optimizers approach the performance of DP. GTDD shows a significant improvement in convergence, achieving the lowest MRC on JOB.

GTDD outperforms RTOS and JOGGER at around 8,000 episodes and achieves performance comparable to DP at around 25,000 episodes, whereas RTOS still results in plans that are 1.6 times more costly than those produced by DP. JOGGER continues to oscillate throughout the training process, ultimately achieving the highest MRC among the compared baselines. Compared to the overall training curve, JOGGER demonstrates more frequent fluctuations. This may be due to the attention tree's inability to retain long-term memory, resulting in the loss of critical information.

Table \ref{tab1:mrc} summarizes the overall performance on both benchmarks. The results indicate that GTDD outperforms other baselines and performs comparably to DP on JOB. On TPC-H, all the methods exhibit similar performance to DP, with the exception of JOGGER, which struggles to handle templates with the largest number of tables.
This disparity can be attributed to the complexity of the dataset: JOB includes 21 tables, with the largest query involving 21 joins, whereas TPC-H includes 8 tables, with the largest query involving 8 joins. Consequently, the primary challenge in TPC-H is not merely the join order, but rather JOB presents greater difficulty in learning the join order due to its larger search spaces.
\begin{table}[t]
\caption{MRC to dynamic programming}
\vspace{-0.1in}
\begin{center}
\begin{tabular}{|c|c|c|c|}
\hline
\textbf{Algorithm} & \textbf{\textit{MRC}} on JOB & \textbf{\textit{MRC}} on TPC-H\\
\hline
\textbf{GTDD} & \textbf{1.06313} & \textbf{1.0000} \\
\hline
\textbf{GTD} & 1.13165 & \textbf{1.0000}\\
\hline
\textbf{RTOS} & 1.22442 & \textbf{1.0000}\\
\hline
\textbf{JOGGER} & 1.54444 & \textbf{1.0000}\\
\hline
\end{tabular}
\label{tab1:mrc}
\end{center}
\vspace{-0.1in}
\end{table}

\subsubsection{Training Time} In this section, we compare the learning efficiency of the proposed method with other learned optimizers. We conducted 40,000 episodes for each approach and recorded the total training time as well as the lowest MRC observed along with the corresponding episode number. We would not evaluate the training time on the latency tuning phase, given that the latency tuning part typically requires a significant amount of time. In order to reduce the latency tuning time, we maintain a latency pool, which stores all the previously observed join orders of each query with the corresponding latency as a key-value pair in a dictionary. To further reduce the tuning time, we utilize the maximum execution time of queries in the training set as a reference point to set a timeout. The timeout is set to $5*max_{latency}(Q)$ which is 10 minutes.
\begin{table}[t]
\caption{Learning efficiency in seconds}
\vspace{-0.1in}
\begin{center}
\begin{tabular}{|c|c|c|c|}
\hline
\textbf{Algorithm} & Optimal & Cost  & Episode\\
\ & MRC & training time & of optimal\\
\hline
\textbf{GTDD} & \textbf{1.06313} & 6323.59742 & 39,000\\
\hline
\textbf{GTD} & 1.13165 & 6185.84266 & \textbf{31,000}\\
\hline
\textbf{RTOS} & 1.22442 & 5868.75733 & 38,000 \\
\hline
\textbf{JOGGER} & 1.54444 &\textbf{5332.49978} & 40,000\\
\hline
\end{tabular}
\label{tab2:trainingtime}
\end{center}
\vspace{-0.1in}
\end{table}

From Table ~\ref{tab2:trainingtime}, we can conclude that GTDD observes the lowest MRC on JOB, indicating that dueling-DQN significantly improves performance compared to vanilla DQN. As expected, GTDD requires the longest training time due to the additional computation overhead associated with Dueling-DQN compared to vanilla DQN. Although GTDD achieves the best MRC at around 39,000 episodes, which is the latest among the other methods, Figure \ref{fig.cost_training} shows that GTDD outperforms GTD, RTOS, and JOGGER after approximately 8,500 episodes and continues to do so for the subsequent 7,500 episodes. GTDD converges at around 22,000 episodes and maintains relatively stable performance, while other baselines continue to exhibit oscillations. JOGGER shows the poorest performance compared to GTDD, GTD, and RTOS. Specifically, GTDD achieves a relatively 6.852\% improvement over GTD, 16.129\% improvement over RTOS, and 48.131\% improvement over JOGGER.
\subsubsection{Latency Tuning}
The ultimate objective of join order selection is to determine the optimal sequence of join orders that minimizes latency. Following the cost training phase, GTDD gains prior knowledge of the data distribution, which enables it to generate low-cost plans. This knowledge can be leveraged to guide the tuning of the pre-trained model to generate plans with lower latency. Building on the model trained with cost, we proceed to fine-tune the model using actual latency to further optimize model performance. 
Figure \ref{fig.latency_tuning} illustrates the results of the latency tuning, demonstrating that most DRL-based models achieve superior solutions compared to DP. In contrast, JOGGER fails to identify better solutions than DP, which is reasonable given that it does not fully leverage prior knowledge during the cost training phase. Notably, GTDD demonstrates a more stable learning process than other baselines. The latency tuning process of GTDD stabilizes after approximately 15,000 episodes, while other baselines continue to exhibit exploration. It is evident that GTDD leverages superior prior knowledge compared to other baselines, leading to a faster and more stable learning process throughout the latency tuning phase.
\begin{table}[t]
\caption{GMRL to dynamic programming}
\begin{center}
\begin{tabular}{|c|c|c|c|}
\hline
\textbf{Algorithm} & \textbf{\textit{GMRL}} on JOB & \textbf{\textit{GMRL}} on TPC-H\\
\hline
\textbf{GTDD} & \textbf{0.60381} & \textbf{0.96399} \\
\hline
\textbf{GTD} & 0.80425 & 0.99225\\
\hline
\textbf{RTOS} & 0.83313 & 1.04111\\
\hline
\textbf{JOGGER} & 1.16758 & 1.05568\\
\hline
\end{tabular}
\label{tab3:gmrl}
\end{center}
\end{table}
Table \ref{tab3:gmrl} shows that GTDD achieves the best GMRL in the latency tuning phase for both JOB and TPC-H. Although GTD has a performance similar to RTOS, Figure \ref{fig.latency_tuning} indicates that the GTD exhibits a more stable training process compared to RTOS, at around 12,000 episodes, GTD becomes relatively stable while RTOS continues to oscillate. 
We further grouped queries by their template ID. Figure \ref{fig.latency_tuning_pattern} demonstrates the GMRL for different templates in JOB. In most cases, GTDD finds better or comparable plans to DP and outperforms other DRL-based models. However, for template \#33, all the DRL-based methods fail to find an execution plan that is optimal and performs drastically poorly. This is because template \#33 is one of the most complex templates, containing 19 join predicates, resulting in a large search space. In templates \#30 and \#32, GTDD finds the join orders close to DP, while other baselines generate worse plans than DP. Although GTDD does not find the best join order for all templates, it still demonstrates superior performance compared to both the other DRL-based optimizers and the native PostgreSQL optimizer. GTDD achieves 0.60381 GMRL on JOB, representing a 20.044\% improvement compared to GTD, a 22.932\% improvement compared to RTOS, and a 56.377\% improvement compared to JOGGER, demonstrating the effectiveness of adapting dueling-DQN over DQN. 

GTDD outperforms RTOS in 87\% of the templates, achieving a significant average improvement of 34.01\%. Conversely, GTDD exhibits slightly inferior performance relative to RTOS in only 13\% of the templates. 
Furthermore, GTDD outperforms JOGGER in 78\% of the templates, yielding an outstanding average improvement of 70.74\%. GTDD shows an extremely slightly inferior performance compared to JOGGER in 22\% of the templates, with a performance regression of 12.09\%.


\begin{figure}[t]
\centerline{\includegraphics[scale=0.3]{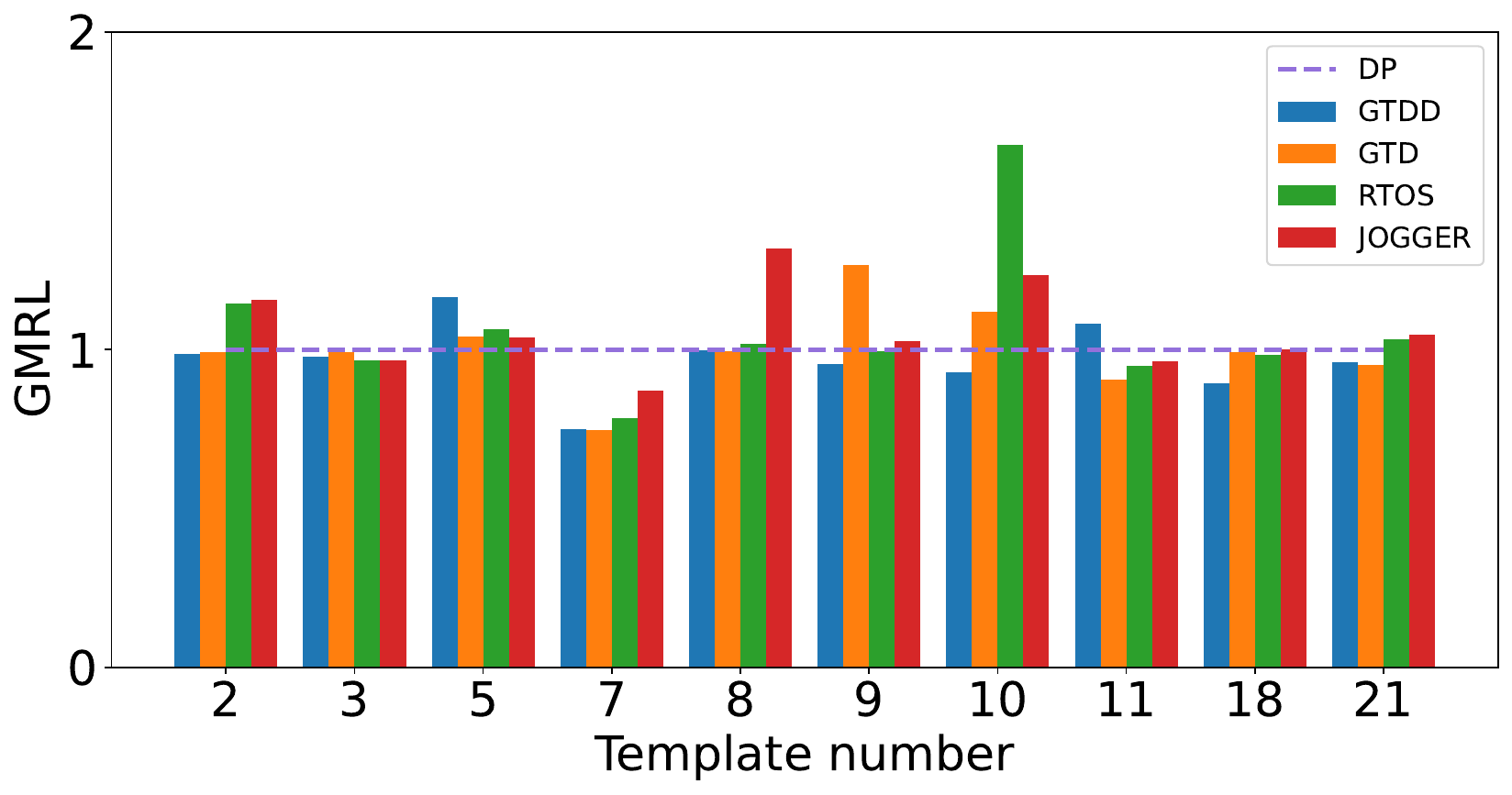}}
\vspace{-0.1in}
\caption{Per-template result using GMRL on TPC-H}
\Description{Result in latency tuning phase based on templates for TPC-H}
\label{fig.latency_tuning_pattern_tpch}
\vspace{-0.1in}
\end{figure}
Figure \ref{fig.latency_tuning_pattern_tpch} demonstrates the performance on TPC-H. We conclude that there's no big difference on this dataset due to a much smaller search spaces. The queries are very simple compared to the queries in JOB, and we can see that GTDD achieves the best results compared to GTD and RTOS. We did not report all 22 templates, as the remaining templates involve only 1 or 2 tables, which means join order selection is not a significant challenge for these queries.
Figure \ref{fig.training_tuning} illustrates the GMRL in both cost training and latency tuning phases 
for JOB. The cost training phase takes 40,000 episodes, followed by the latency tuning phase, which consists of 20,000 episodes. Initially, the model's performance is poor while it is exploring, resulting in disastrous plans. At 13,000 episodes, the GMRL of GTDD fluctuates around 1 until the latency tuning phase begins, indicating that it has sufficiently learned the underlying data distribution with estimated costs. Upon transitioning to the latency tuning phase, the GMRL of GTDD begins to change and initially performs slightly worse than DP as it explores the environment with new feedback (e.g. latency). Eventually, GTDD surpasses DP in finding better plans with latency as feedback and ultimately converges to a GMRL below 1.
\section{Conclusion}
In this work, we propose a novel framework, GTDD, which tackles holistically the problem of join order selection, by accurately representing the input data and leveraging the full potential of reinforcement learning. Specifically, it employs representation learning for representing queries at different levels, and dueling-DQN to tackle the limitation of the unstable training process in DRL. GTDD uses Tree-LSTM to capture the sequence of the join order and the intermediate join plan. Furthermore, GTDD uses a novel reward function. We conduct a series of experiments on two benchmarks and demonstrate that GTDD has significantly better results and a more stable training process compared to the state-of-the-art DRL-based techniques that perform join order selection.
\bibliographystyle{ACM-Reference-Format}
\bibliography{main}
\end{document}